\documentclass[english]{revtex4}
\usepackage[T1]{fontenc}
\usepackage[latin1]{inputenc}
\usepackage{float}
\usepackage{amsmath}
\usepackage{graphicx}
\usepackage{amssymb}

\makeatletter
\usepackage{babel}
\makeatother
\begin{document}

\title{Probing the intrinsic shot noise of a Luttinger Liquid through impedance
matching.}

\author{K.-V. Pham}

\affiliation{Laboratoire de Physique des Solides, Université Paris-Sud, 91405
Orsay, France.}

\begin{abstract}
We argue that a simple way to bypass reflections at the boundaries
of a finite Luttinger liquid (LL) connected to electrodes is to match
load and drain impedances to the characteristic impedance of the LL
viewed as a mesoscopic transmission line.

For an impedance matched LL, this implies that the AC and DC shot
noise properties of a finite LL are identical to those of an infinite
LL.

Even for an impedance mismatched LL, we show by a careful analysis
of reflections that the intrinsic infinite LL properties can still
be extracted yielding possibly irrational charges for the LL elementary
excitations. We improve on existing results for AC shot noise by deriving
expressions with explicit dependence on the charges of the fractional
states. Most notably these results can be established quite straightforwardly
without resort to the Keldysh technique.

We apply these arguments to two experimental setups which allow the
observation of different sets of fractional quasiparticles: (i) injection
of current by a STM tip in the bulk of a LL; (ii) backscattering of
current by an impurity. 
\end{abstract}
\maketitle

\section{Introduction.}

Shot noise is a topic of current interest because it allows access
to non-equilibrium transport properties of a system and notably to
the charge carried by the elementary excitations \cite{noise}. For
a standard non-disordered Fermi liquid shot noise yields a unit charge
for the Landau quasiparticle but in the Fractional Quantum Hall Effect
(FQHE) shot noise has revealed rational charges for the famous Laughlin
quasiparticles \cite{fqhe}.

The Luttinger liquid \cite{ll} is another example of strongly-correlated
system where elementary excitations with non-integral charges are
expected. While the standard bosonization picture of the LL stresses
plasmon-like excitations and zero modes \cite{bosonisation}, that
picture is unconvenient to interpret shot noise because the charged
excitations have no dynamics (they are zero-modes with no dispersion);
an alternative 'fractional states picture' of the charged excitations
was recently developped \cite{fractionalization}: it was shown that
there are other bases of exact eigenstates for the LL consisting of
states carrying in general irrational charges (a summary will be found
in Appendix A). The fractional states are created in pairs with a
total charge which is always an integer.

These fractional eigenstates permit a straight interpretration of
earlier shot noise results for an infinite LL with an impurity, where
a charge $K$ was found in the shot noise \cite{kane} ($K$ is the
usual LL parameter). A very recent calculation for the shot noise
due to current injection by a STM tip in an infinite LL also found
that charges $\frac{1+K}{2}$and $\frac{1-K}{2}$ are involved \cite{stm}.
States with such unconventional charges are difficult to account with
in the standard bosonized picture of the LL while they come out naturally
in the 'fractional states picture' of the LL, where they had earlier
been predicted and built as exact eigenstates of the LL hamiltonian
\cite{fractionalization}.

In spite of these theoretical results for the shot noise great strides
are still needed toward an experimental verification for the following
reason: the experimental systems have a finite length and are mostly
at the mesoscopic scale, which makes it impossible to ignore the influence
of electrodes on the transport properties of the LL. Theorists have
therefore been busy trying to model reservoirs and compute conductance
as well as current fluctuations for the LL: it is now accepted that
the conductance of the finite LL connected to two electrodes differs
from the conductance of the infinite LL \cite{ill}. Shot noise has
also been computed by several groups for the following setups: (i)
STM tip injecting current \cite{stm}; (ii) backscattering due to
an impurity \cite{impurity,pono}, and claims have been made that
charges of fractional excitations can be recovered from AC shot noise.

However several criticisms can be addressed to those claims:

- (a) the calculations depend on a very specific model, the inhomogeneous
LL which models electrodes as 1D free fermions \cite{ill}, which
is a debatable assumption; to what extent these calculations are sufficiently
general and model-independent is hard to assess.

- (b) Both the charge of fractional excitations and the reflection
parameter used in these models depend on the LL parameter $K$ : but
the Fano factor mixes contributions from both the fractional charge
and the reflection parameter; this implies that a measurement of the
Fano factor reduces to just measuring the LL parameter $K$ and is
not a direct measurement of the charge of the charge-carriers (contrast
eq.(\ref{eq:stm-noise}) with our results, e.g. eq.(\ref{eq:stm-new})
). In the case of an impurity in a LL the AC Fano factor $F(\omega)$
is a periodic function with period $\omega_{L}=\frac{\pi u}{L}$ (for
the unlikely case of an impurity sitting exactly in the middle of
the LL) and it was claimed \cite{impurity} that averaging over one
period yields the fractional charge $K$: it is difficult to understand
why it should be so and actually as we will show in this paper this
result is a model-dependent accident and in general averaging over
a period does NOT yield the charge of the fractional excitation of
the LL.

We use in this paper a general formalism encompassing the inhomogeneous
LL and can thus address these qualms:

- (a) we indeed showed in several earlier papers that the inhomogeneous
LL is actually curtailed to very specific experimental conditions
(interface resistances of the LL to the electrodes pinned at half-a-quantum
of resistance $R_{0}=\frac{h}{2e^{2}}$): it is a special case of
our theory, which is itself valid for arbitrary values of the interface
resistances at the electrodes \cite{p3}.

- (b) our theory for a LL in a multi-terminal environment has the
advantage of being able to sort out from the Fano factor the contributions
coming from reflections and those coming from the fractional charges.
As a result it does show unambiguously how to get access to the fractional
charges. 

To perform shot noise calculations a tool of choice is of course the
Keldysh technique. That approach can also be used within our framework
(and is sketched in Appendix \ref{sec:Green's-functions-using}).
However in the finite geometry in the course of the Keldysh calculation
the distinction between the dual role of the LL parameter $K$ (which
intervenes both in the charges scattered and in the reflection coefficients)
is blurred so that the difficulties explained above apply (point (b)
). We follow another route in this paper and show the essential physics
(and almost all of the earlier results found with the inhomogeneous
LL) can be retrieved from a fine analysis of reflections and from
viewing the LL as a mesoscopic transmission line. 

More importantly such an approach yields insights on how to bypass
the effect of reflections, which (although yielding an interesting
physics in themselves), are largely a nuisance as far as measuring
the fractional charges is concerned: to avoid reflections in a transmission
line it is sufficient to match the load and drain impedances to the
characteristic impedance of the line. As a result the system behaves
as if it were an infinite system. The AC shot noise properties of
an impedance matched LL are therefore exactly those of an infinite
system.

The outline of the paper will be as follows:

- Section II is centered on the concept of impedance matching for
a LL. It will introduce notations, explain the formalism used to describe
the LL connected to reservoirs; that formalism makes use of boundary
conditions describing quite straightforwardly the coupling of a LL
to interface resistances. This assumes ohmic coupling of the LL to
electrodes. The section concludes with the proof that the impedance
matched LL is equivalent to an infinite LL.

- Section III discusses the setup of a STM tip injecting electrons
in a LL and generalizes results found within the framework of the
inhomogeneous LL.

- Section IV shows how to use calculations done for the STM tip setup
to infer AC shot noise for another apparently unrelated setup, that
of an impurity which backscatters current in a LL. We again generalize
the inhomogeneous LL results.

\section{Impedance matching.}

\subsection{LL as a mesoscopic transmission line.}

(For earlier discussions of the LL as a $LC$ line the reader is refered
for example to Ref. \cite{lc,p3}).

The LL is just a quantum \emph{LC} line as evidenced from its hamiltonian:
\[
H=\int_{-a}^{a}dx\,\,\frac{hu}{4K}\rho^{2}+\frac{hu\, K}{4}\overline{j}^{2}\]
 where $\overline{j}=\rho_{+}^{0}-\rho_{-}^{0}$ is the difference
between bare right and left electron densities (at right and left
Fermi points $\pm k_{F}$). Rewriting the Hamiltonian in terms of
charge density and current: \[
\rho_{e}=e\,\,\rho;\; j_{e}=e\,\, u\, K\overline{\, j}\]
 (the last expression follows from charge conservation and the equations
of motion) there follows: \[
H=\int_{-a}^{a}dx\,\,\frac{hu}{4Ke^{2}}\rho_{e}^{2}+\frac{h}{4u\, Ke^{2}}j_{e}^{2}\]
 which shows the LL has both capacitance and inductance per unit length:\[
\mathcal{C=}\frac{2Ke^{2}}{h\, u},\quad\mathcal{L}=\frac{h}{2u\,\, Ke^{2}}.\]
 These capacitive and \emph{inductive} behaviours show up quite well
in the LL dynamical conductance and dynamical impedance as computed
in Ref. \cite{p3}.

\subsection{Basics of transmission lines: Impedance mismatch.}

As for any finite transmission line (or for that matter, any sound
wave in a tube, etc...) for the open system (rigid boundaries) one
has standing waves due to perfect reflection at the boundaries.

Such a transmission line is characterized by a characteristic impedance
\begin{equation}
Z_{0}=\frac{h}{2Ke^{2}}=\sqrt{\mathcal{L}/\mathcal{C}}.\label{z0}\end{equation}
 We remind the reader that a characteristic impedance IS NOT a standard
DC or AC impedance: it is rather an instantaneous or \textit{surge
impedance} seen by the electrical wave as it moves along the \emph{LC}
line. The stark difference can be seen for example for a resistanceless
\emph{LC} line: the DC impedance is then zero while the characteristic
impedance is non-zero. For the usual \emph{LC} line $Z_{0}$ obeys:
\[
i^{+}(x,t)=Z_{0}\, V^{+}(x,t),\quad i^{-}(x,t)=-Z_{0}\, V^{-}(x,t)\]
 where the total voltage signal is $V=V^{+}+V^{-}$ and $i^{\pm}$
are right (or left) moving currents so that $i=i^{+}-i^{-}$.

If one now attaches load impedances $Z_{S}$ and $Z_{D}$ at the interfaces
of a length $L$ transmission line such that: \begin{equation}
\frac{i(0)}{V(0)}=Z_{S},\quad\frac{i(L)}{V(L)}=Z_{D}\label{impedance}\end{equation}
 freshman physics tells us also that there will be reflection at the
boundaries $x=0$ and $x=L$ with respectively reflexion coefficients
(for the plasma wave): \begin{equation}
r_{S}=\frac{Z_{S}-Z_{0}}{Z_{S}+Z_{0}},\quad r_{D}=\frac{Z_{D}-Z_{0}}{Z_{D}+Z_{0}}\text{.}\label{reflec}\end{equation}
 This implies that reflexions are killed whenever the load impedances
at the source and drain are equal to the characteristic impedance.
This leads to the concept of \textbf{impedance matching} well-known
to electrical engineers: by carefully matching the load impedances
one eliminates undesired reflections and one can thus have a \textit{maximal
energy transfer} from source to load.

Furthermore for load impedances attached to a resistanceless transmission
line one has a series addition law and therefore: \begin{equation}
G=\frac{1}{Z_{S}+Z_{D}}.\label{conductance}\end{equation}

\subsection{Formalism used in this paper: 'impedance boundary conditions'.}

We will use a description of the LL connected to reservoirs \cite{p3}
which is the exact implementation of the customary load impedances
boundary conditions described in the previous subsection and used
for transmission lines or sound waves in tubes:

\begin{eqnarray}
i(-a,t) & = & \frac{1}{Z_{S}}\left(V_{S}(t)-\frac{\delta H}{\delta\rho(-a,t)}\right)\label{bound1}\\
i(a,t) & = & \frac{1}{Z_{D}}\left(\frac{\delta H}{\delta\rho(a,t)}-V_{D}(t)\right)\end{eqnarray}

\begin{figure}
\includegraphics{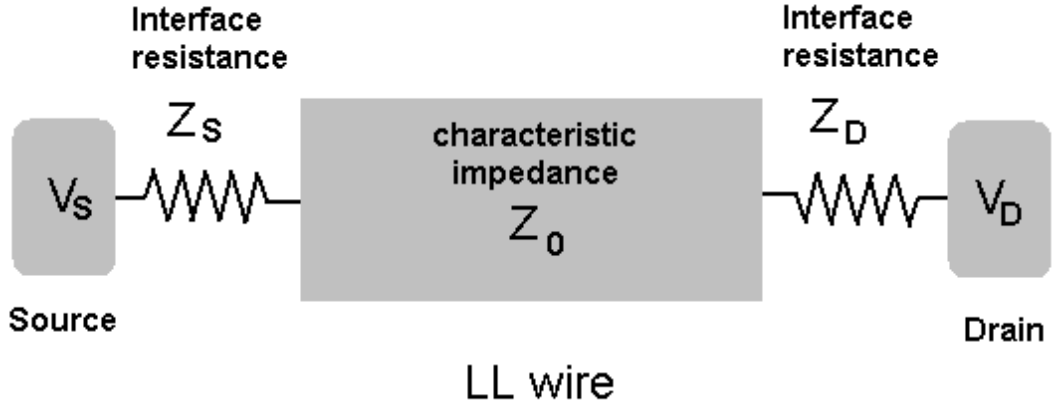}

\caption{\label{fig-formalisme}Impedance boundary conditions: the LL wire
is connected to two electrodes at voltages $V_{S}$ and $V_{D}$ through
two boundary impedances.}
\end{figure}

$Z_{S}$ and $Z_{D}$ are interface impedances at respectively the
source and the drain \textit{which for simplicity will be assumed
to be real numbers} throughout the paper (but more general situations
could be discussed), $i(x,t)$ is the current operator, and source
and drain are set at a voltage $V_{S}$ or $V_{D}$ (see Fig.\ref{fig-formalisme}).
The Heisenberg picture is assumed so that we work with time-dependent
operators. Since $\frac{\delta H}{\delta\rho(x,t)}$ is the energy
needed to add locally a particle, it corresponds to a local chemical
potential for the LL. Actually the boundary conditions are tantamount
to assuming Ohm's law at the boundaries of the system: the current
is proportional to a voltage drop between the reservoir and the LL
wire and the proportionality constant is just an interface resistance.

For calculations it is convenient to introduce chiral chemical potentials
corresponding to the chiral plasmons of the LL.

We consider the standard Luttinger Hamiltonian for a wire of length
$L=2a$. \begin{equation}
H=\int_{-a}^{a}dx\,\,\frac{hu}{2K}\left(\rho_{+}^{2}+\rho_{-}^{2}\right)\end{equation}

$\rho_{+}$ and $\rho_{-}$ are chiral particle densities which obey
the relation $\rho_{\pm}(x,t)=\rho_{\pm}(x\mp ut)$. Their sum is
just the total particle density $\rho-\rho_{0}$ while the electrical
current is simply $i(x,t)=eu\left(\rho_{+}-\rho_{-}\right)$.

We now define the following operators:\begin{equation}
\mu_{\pm}(x,t)=\frac{\delta H}{\delta\rho_{\pm}(x,t)}\end{equation}
 Physically they correspond to chemical potential operators: their
average value yields the energy needed to add one particle at position
$x$ to the chiral density: $\rho_{\pm}\longrightarrow\rho_{\pm}+\delta(x)$.
These chiral chemical potentials correspond to the plasma chiral eigenmodes
of the Luttinger liquid and \textit{not to the left or right moving
(bare) electrons}.

From their definition it follows that: \begin{equation}
\mu_{\pm}(x,t)=\frac{hu}{K}\rho_{\pm}(x,t),\end{equation}
 and therefore using the definition eq.(\ref{z0}):

\begin{eqnarray}
i(x,t) & =K\frac{e}{h}\left(\mu_{+}(x,t)-\mu_{-}(x,t)\right)\nonumber \\
 & =\frac{1}{2eZ_{0}}\left(\mu_{+}(x,t)-\mu_{-}(x,t)\right)\label{currentb}\end{eqnarray}
 Since these operators have a chiral time evolution: \begin{equation}
\mu_{\pm}(x,t)=\mu_{\pm}(x\mp ut),\end{equation}
 it follows also: \begin{eqnarray}
\mu_{+}(a,\omega) & =\exp i\phi\,\,\mu_{+}(-a,\omega),\\
\mu_{-}(a,\omega) & =\exp-i\phi\,\,\mu_{-}(-a,\omega),\end{eqnarray}
 where we have defined a plasmon phase $\phi$ accumulated along the
wire as:\begin{equation}
\phi=\omega\frac{2a}{u}.\end{equation}

\subsection{DC conductance of the impedance mismatched system.}

This section is mostly present for pedagogical reasons. We first review
the derivation of the DC conductance in this formalism (earlier discussion
can be found in \cite{p1} and \cite{p2}). We will then show how
the calculation is interpreted in terms of reflections of the plasma
wave; this provides a simple context which will help us later when
we seek to interpret this paper's results on the shot noise.

\subsubsection{A straight derivation.}

The boundary conditions can be rewritten in terms of the chiral chemical
potentials as: \begin{eqnarray}
Z_{S}i(-a,t) & = & \left(V_{S}(t)-\frac{\mu_{+}(-a,t)+\mu_{-}(-a,t)}{2e}\right),\nonumber \\
Z_{D}i(a,t) & = & \left(\frac{\mu_{+}(a,t)+\mu_{-}(a,t)}{2e}-V_{D}(t)\right).\label{bc}\end{eqnarray}

In the DC regime these operators lose their space and time-dependence
therefore adding the two previous equations yield: $(Z_{S}+Z_{D})\, i=V_{S}-V_{D}$.
The DC conductance is thus: \[
G=\frac{1}{Z_{S}+Z_{D}},\]
as expected for a resistanceless LC-line connected in series to two
load impedances. For the impedance matched system:\[
Z_{S}=Z_{0}=Z_{D}\]
and therefore using eq.(\ref{z0}): \[
G=\frac{1}{Z_{0}}=\frac{Ke^{2}}{h},\]
which is exactly the DC conductance of the infinite system \cite{kcond}.

The same considerations can be applied to the AC conductance matrix
\cite{p3} and one finds that the impedance-matched system has exactly
the properties of the infinite system.

\subsubsection{Physical interpretation: an equivalent derivation using reflections.}

As seen in the previous subsection the computation of the DC conductance
is quite straightforward using the 'impedance boundary conditions'.
For the sake of pedagogy we will rederive the DC conductance as a
function of \textit{reflection coefficients}.

We consider for full generality arbitrary reflections coefficients
at the source and drain $r_{S}$ and $r_{D}$. The source and drain
are set at voltages $V_{S}$ and $V_{D}$. We now build the contributions
to the current resulting from the multiple reflections.

\textbf{Order zero:}we start with the values in the infinite system
for the chiral currents injected by source and drain (see for instance
\cite{key-5}) as resulting from a straightforward linear-response
calculation: \[
i_{0}^{+}=\frac{V_{S}}{2Z_{0}},\qquad i_{0}^{-}=\frac{V_{D}}{2Z_{0}}\]
 which yields $I=i_{0}^{+}-i_{0}^{-}=\frac{V_{S}-V_{D}}{2Z_{0}}=\frac{Ke^{2}}{h}\left(V_{S}-V_{D}\right)$.
Resulting for the infinite system to a conductance renormalized by
interactions: $G=KG_{0}$.The $\pm$ exponent corresponds to the chirality
(right or left moving plasmons).

\textbf{Order one:} we take into account the reflections at the boundaries.
This implies additional currents:\[
i_{1}^{+}=r_{S}~i_{0}^{-}-r_{S}~i_{0}^{+},\qquad i_{1}^{-}=r_{D}~i_{0}^{+}-r_{D}~i_{0}^{-}\]
 Each current has two contributions at this order: the first one corresponds
to the reflexion at the boundary of the current of opposite chirality
while the second one takes into account the fact that a fraction of
the incoming current does not actually enter the system due to reflexion.

\begin{figure}[t]
\includegraphics{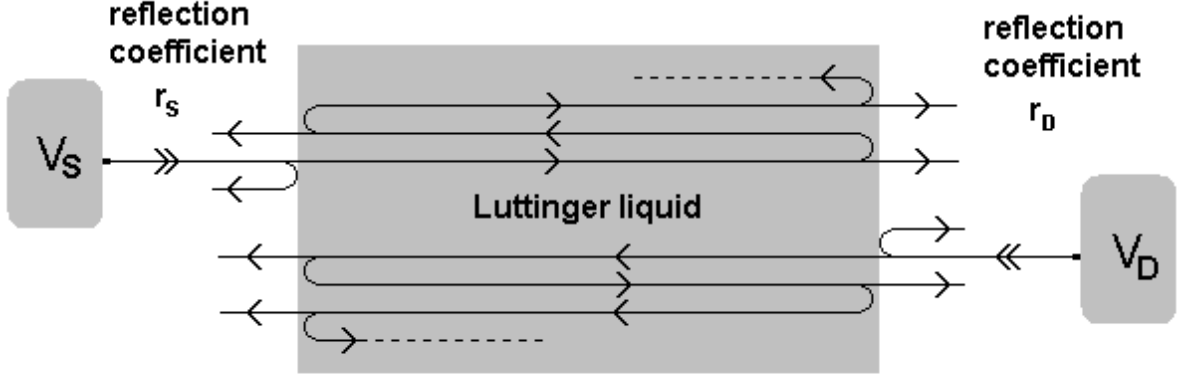}

\caption{Multiple reflections renormalizing the conductance.}
\end{figure}

\textbf{Order} $n\geq2$:

We have multiple reflexions of the chiral currents within the system.
Therefore one simply has: \[
i_{n}^{+}=r_{S}~i_{n-1}^{-},\qquad i_{n}^{-}=r_{D}~i_{n-1}^{+}\]
 Let us now collect each contribution to get the total current.\begin{align*}
\left(\begin{array}{c}
i^{+}\\
i^{-}\end{array}\right) & =\left(\begin{array}{c}
i_{0}^{+}\\
i_{0}^{-}\end{array}\right)+\sum_{n=0}^{\infty}\left(\begin{array}{cc}
0 & r_{S}\\
r_{D} & 0\end{array}\right)^{n}\left(\begin{array}{c}
i_{1}^{+}\\
i_{1}^{-}\end{array}\right)\\
 & =\left(\begin{array}{c}
i_{0}^{+}\\
i_{0}^{-}\end{array}\right)+\frac{1}{1-r_{S}r_{D}}\left(\begin{array}{cc}
1 & r_{S}\\
r_{D} & 1\end{array}\right)\left(\begin{array}{cc}
-r_{S} & r_{S}\\
r_{D} & -r_{D}\end{array}\right)\left(\begin{array}{c}
i_{0}^{+}\\
i_{0}^{-}\end{array}\right)\\
 & =\frac{1}{1-r_{S}r_{D}}\left(\begin{array}{cc}
1-r_{S} & r_{S}-r_{S}r_{D}\\
r_{D}-r_{S}r_{D} & 1-r_{D}\end{array}\right)\left(\begin{array}{c}
i_{0}^{+}\\
i_{0}^{-}\end{array}\right)\end{align*}

Thus: \[
I=i^{+}-i^{-}=\frac{1}{1-r_{S}r_{D}}\left(1-r_{S}-r_{D}+r_{S}r_{D}\right)\frac{V_{S}-V_{D}}{2Z_{0}}.\]

and the conductance is: \begin{equation}
G=\frac{1-r_{S}-r_{D}+r_{S}r_{D}}{1-r_{S}r_{D}}\frac{1}{2Z_{0}}.\label{cond}\end{equation}

What do we learn from this calculation?

- (i) the multiple reflections are indeed not innocuous: they are
at the heart of the renormalization of the conductance.

- (ii) for arbitrary values of the reflection coefficients the conductance
can take any value.

- (iii) for reflection coefficients equal to zero one recovers the
infinite system physics.

The third point may sound like a tautology but to radiowave engineers
used to transmission lines this is but a statement of \textbf{impedance
matching}, a concept whose importance has yet been unrecognized for
the LL and the central issue of this paper. 

-(iv) Let's make contact with the 'impedance boundary conditions',
which can be rewritten as:\begin{eqnarray*}
Z_{S}ue\left(\rho_{+}(-a)-\rho_{-}(-a)\right) & = & V_{S}-Z_{0}ue\left(\rho_{+}(-a)+\rho_{-}(-a)\right),\\
Z_{D}ue\left(\rho_{+}(a)-\rho_{-}(a)\right) & = & Z_{0}ue\left(\rho_{+}(a)+\rho_{-}(a)\right)-V_{D}.\end{eqnarray*}
The values of the reflection coefficients can then be recovered; of
course as expected from the analogy to a classical LC-transmission
line one finds:\[
r_{S}=\frac{Z_{S}-Z_{0}}{Z_{S}+Z_{0}},\quad r_{D}=\frac{Z_{D}-Z_{0}}{Z_{D}+Z_{0}}\text{.}\]
This is unsurprising given that the LL hamiltonian is quadratic and
that therefore the classical equations of motion are exact.

Plugging in these values of the reflection coefficients into the expression
of the conductance above ( eq.(\ref{cond}) ) one recovers as it should
be:\[
G=\frac{1}{Z_{S}+Z_{D}}.\]

\subsection{Relation to the inhomogeneous LL and other models of a LL connected
to leads.}

The significance of reflections for a LL connected to leads was first
recognized using the inhomogeneous LL, which is a model using a space
dependent LL parameter: $K(x)=K$ for $\left|x\right|<L/2$ and $K(x)=1$
for $\left|x\right|\geq L/2$ for Fermi liquid leads \cite{ill}.

It can be shown (see Appendix of Ref. \cite{p3}) that the inhomogeneous
LL and several other theories based on boundary conditions (such as
the 'radiative boundary conditions' \cite{key-6} ) actually obeys
our 'impedance boundary conditions' albeit in a very specific case,
when load and drain impedances take the values:\[
Z_{S}=Z_{D}=\frac{h}{2e^{2}}.\]

These earlier theories will therefore be valid only for rather clean
contacts with impedances close to those of a non-interacting system.
Our more general approach has the advantage of not making such assumptions.

\subsection{Proof that the impedance matched LL is equivalent to an infinite
LL.}

The proof is simple: the impedance matched LL is equivalent to an
infinite LL because the Green's function of a LL subjected to the
'impedance boundary conditions' is identical with that of an infinite
LL \emph{provided} $Z_{S}=Z_{D}=Z_{0}$ (impedance matching condition).

The Green's function will not be used in this paper and the reader
will find details in Appendix \ref{sec:Green's-functions-using} which
sketches its derivation. This is the starting point for a Keldysh
treatment using the 'impedance boundary conditions'.

\section{Injection of particles through a STM tip.}

In this section we discuss the following setup: a STM tip tunneling
electrons into the bulk of a LL.

The obvious strategy to tackle the transport properties is to use
the Keldysh formalism, which is well suited to non-equilibrium physics:
this is the approach followed by Crépieux, Martin et coll., \cite{stm}
using the inhomogeneous LL. Actually the same can be done with the
'impedance boundary conditions' ; the main difference is that the
Green functions of the inhomogeneous LL correspond to the special
case $Z_{S}=Z_{D}=\frac{h}{2e^{2}}$ and that one has two reflection
coefficients (at source and drain) instead of a single one. That approach
is sketched in APPENDIX \ref{sec:Green's-functions-using}. 

We will show another more economical and physically more transparent
approach which yields most of the physics (and often gives the same
results). It is based upon making a distinction between operators
for the injected currents and operators for the measured current.

We discuss separately DC and AC shot noise because our approach is
simpler to understand in the DC case; additionally:

- for the DC noise, we discuss both asymetric injection of particles
(unequal probability to inject an electron either at the left or the
right Fermi point) and arbitrary interface resistances (in contrast
Ref. \cite{stm} deals with symetric injection and implies interface
resistances set to $\frac{h}{2e^{2}}$).

- for AC noise while keeping arbitrary interface resistances (which
is the required setting for a discussion of impedance matching) we
restrict for simplicity to symetric injection.

We also restrict ourselves in what follows to 'excess noise' and never
discuss 'equilibrium noise' since the latter is in some sense trivial
as it obeys the fluctuation-dissipation theorem.

\subsection{DC shot noise.}

Our derivation of the DC shot noise wil follow these steps:

- we first relate the current operators to another set of operators
('injected curent operators')(section \ref{sub:Injected-currents-versus});

- we compute the excess noise of these operators (section \ref{sub:Excess-noise-for-injected});

- and finally infer from them the current excess noise (section \ref{sub:Excess-noise-for-measured}).

\subsubsection{Earlier results}

We consider the following apparatus: an STM tip tunnels electrons
to the bulk of a LL (say, a carbon nanotube). 

\begin{figure}
\includegraphics{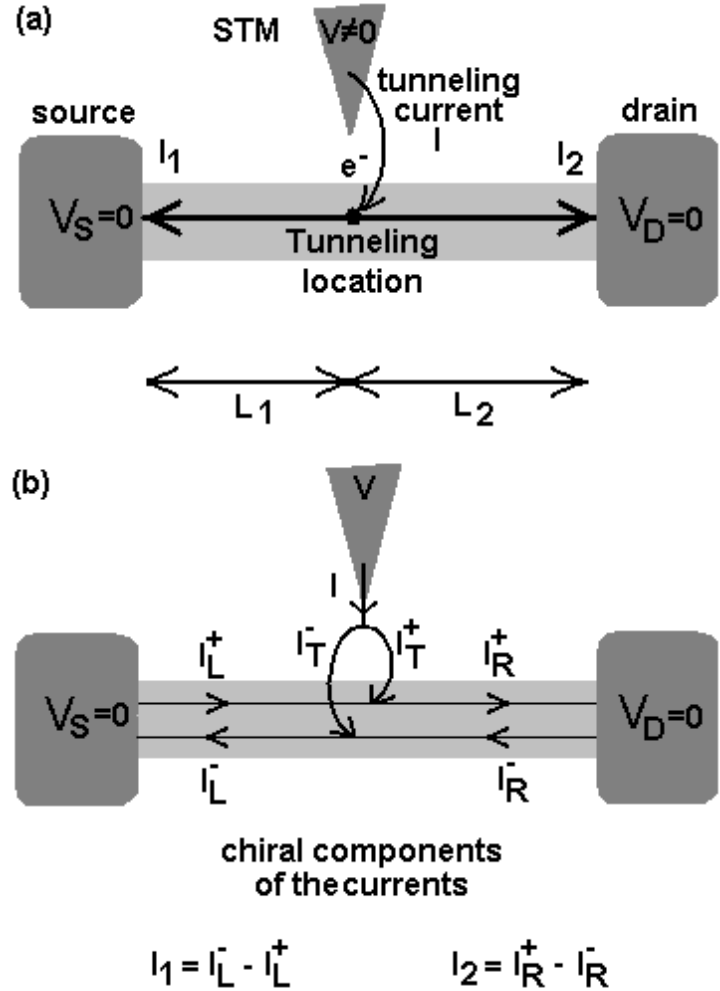}

\caption{\label{cap:STM-tip-injects}STM tip injects electrons into a LL.}
\end{figure}

We call $I_{1}$ and $I_{2}$ the currents going to the left and to
the right of the injection point. These currents are oriented OUTgoing
from the injection point. They are the currents measured respectively
at the source and drain. The main work on the subject is that of Crépieux,
Martin and coll \cite{stm}. They find that in the infinite system
the direct correlations of current and cross-correlations obey:

\begin{align*}
F_{1}^{\infty}=\frac{\left\langle \Delta I_{1}^{2}\right\rangle }{\left\langle \Delta I_{1}\right\rangle } & =\left(Q_{+}^{2}+Q_{-}^{2}\right)\\
F_{2}^{\infty}=\frac{\left\langle \Delta I_{2}^{2}\right\rangle }{\left\langle \Delta I_{2}\right\rangle } & =\left(Q_{+}^{2}+Q_{-}^{2}\right)\\
\frac{\left\langle \Delta I_{1}\Delta I_{2}\right\rangle }{\left\langle \Delta I_{1}\right\rangle } & =2Q_{+}Q_{-}\end{align*}

where $Q_{\pm}=\frac{1\pm K}{2}$ are the (in general irrational)
charges of two fractional states: injection of a $+k_{F}$ electron
was proved to result in the creation of two exact fractional eigenstates
of the LL \cite{fractionalization} propagating in the right and left
directions and carrying just such charges $Q_{\pm}=\frac{1\pm K}{2}$.
(Injection of a $-k_{F}$ electron results in the opposite: propagation
to the left of charge $\frac{1+K}{2}$ and propagation to the right
of charge $\frac{1-K}{2}$). These peculiar states are a combination
of one Laughlin quasiparticle with one holon. A noteworthy observation
is that Cr\'{e}pieux et al \cite{stm}, find POSITIVE cross-correlations
which is quite unexpected for a fermionic system. 

(As an aside we note that such charges were anticipated in \cite{key-3},
where \emph{as a quantum average} a charge density $e$ $\left\langle \rho(x,t=0)\right\rangle =e\,\delta(x)$
was found to separate into two charge packets $\left\langle \rho(x,t)\right\rangle =\frac{1+K}{2}e\,\delta(x-ut)+\frac{1-K}{2}e\,\delta(x+ut)$
carrying exactly the charges $\frac{1\pm K}{2}$; it was realized
later in \cite{fractionalization} that these charges are actually
carried by \emph{exact fractional eigenstates}).

Lebedev et al \cite{stm} later found that these results were invalidated
when the LL is connected to Fermi liquid reservoirs; one gets up to
order two in perturbation: \begin{align*}
F_{1}=\frac{\left\langle \Delta I_{1}^{2}\right\rangle }{\left\langle \Delta I_{1}\right\rangle } & =e\\
F_{2}=\frac{\left\langle \Delta I_{2}^{2}\right\rangle }{\left\langle \Delta I_{2}\right\rangle } & =e\\
\frac{\left\langle \Delta I_{1}\Delta I_{2}\right\rangle }{\left\langle \Delta I_{1}\right\rangle } & =0.\end{align*}
 So one recovers integral charges. This disappointing result is explained
by Lebedev et al. as follows \cite{stm}: in a second-order perturbation
theory one neglects correlation between the transport of two electrons
injected sequentially; assuming perfect transmission of the electron,
a single injected electron will be transmitted as a whole to either
one of the reservoirs so that $\left\langle Q_{1}Q_{2}\right\rangle =0$
which results into $\left\langle \Delta I_{1}\Delta I_{2}\right\rangle =0$.
As can one can see a crucial element of such an argumentation is \emph{perfect
transmission} of the injected electron. We will show in the course
of this paper that such a condition can be relaxed (it actually depends
crucially on the impedances at the boundaries of the system) which
results in different values of the current-current correlators predicted
by Crépieux, Martin and coll. \cite{stm}. 

(NB: As a shorthand notation we have written in the above the current
correlation $\left\langle \Delta I_{1}^{2}\right\rangle $ for the
zero frequency Fourier transform $\left\langle \Delta I_{1}^{2}(\omega=0)\right\rangle =\left\langle \int dt\,\Delta I_{1}(t)\,\Delta I_{1}(0)\right\rangle $
of that quantity. We will also use this notation in the rest of the
paper.)

\subsubsection{Injected currents versus measured currents\label{sub:Injected-currents-versus}.}

(We work here in the Heisenberg picture for the operators. Since we
consider actually in this subsection a DC context the time dependence
will be dropped.)

The goal of that subsection is to show that the operators for the
measured currents $I_{1}$ and $I_{2}$ are NOT the operators for
the currents $I_{T}^{-}$ and $I_{T}^{+}$ injected by the STM tip.
The reason for that is simple: there are reflections at the boundaries.
Imagine for instance that the STM tip injects current only to the
left of the tip. Due to reflections eventually there must be some
current flowing to the right which shows that the net currents $I_{1}$
and $I_{2}$ flowing in the system differ from the currents injected
by the tip.

The rationale for making such distinctions is that the noise properties
of $I_{T}^{-}$ and $I_{T}^{+}$ can be easily found and that from
them the correlators for $I_{1}$ and $I_{2}$ can then be inferred
very simply without recourse to the Keldysh technique.

Let us see that in detail. 

Firstly we note that in this DC context the operators will have no
space or time dependence; but due to the presence of the tunneling
point we must distinguish the values of the operators to the left
or to the right of the STM tip. Accordingly we note:\begin{eqnarray*}
\mu_{R}^{\pm} & = & \mu^{\pm}(x>0),\:\mu_{L}^{\pm}=\mu^{\pm}(x<0);\\
i_{R}^{\pm} & = & i^{\pm}(x>0),\: i_{L}^{\pm}=i^{\pm}(x<0);\\
I_{2} & = & I(x>0)\\
I_{1} & =- & I(x<0)\end{eqnarray*}

As we can see from Fig. \ref{cap:STM-tip-injects} the operators for
the current flowing to the left and to the right of the STM tip and
the operators for the currents injected are different; by definition
and using eq.(\ref{currentb}), \begin{align*}
I_{2} & =\frac{1}{2eZ_{0}}\left(\mu_{R}^{+}-\mu_{R}^{-}\right)=i_{R}^{+}-i_{R}^{-}\\
-I_{1} & =\frac{1}{2eZ_{0}}\left(\mu_{L}^{+}-\mu_{L}^{-}\right)=i_{L}^{+}-i_{L}^{-}\end{align*}
 where: \[
i_{R/L}^{\pm}=\frac{\mu_{R/L}^{\pm}}{2eZ_{0}}\]
 is the current carried by each chiral branch on the right (index
$R$) or the left (index $L$) of the tunnelling tip (see Fig.2 (b)
). In other words what we call $I_{1}$ and $I_{2}$are just the currents
flowing in the wire. 

In contrast the operators for the currents injected by the STM tip
to its right and to its left are respectively:

\begin{align*}
I_{T}^{+} & =\frac{1}{2eZ_{0}}\left(\mu_{R}^{+}-\mu_{L}^{+}\right)=i_{R}^{+}-i_{L}^{+}\\
I_{T}^{-} & =\frac{1}{2eZ_{0}}\left(\mu_{L}^{-}-\mu_{R}^{-}\right)=i_{L}^{-}-i_{R}^{-}\end{align*}
 The reason for that is that the eigenmodes of the LL are NOT left
or right moving electrons: the current injected by the tip really
goes into the chiral (plasmon) eigenmodes of the LL.

In the infinite system that discrepancy between injected and measured
currents is irrelevant (for DC measurements) because for a LL wire
which is grounded obviously $\left\langle i_{L}^{+}\right\rangle =0=\left\langle i_{R}^{-}\right\rangle $
since no current is coming from the electrodes: as a result $\left\langle I_{1}\right\rangle =\left\langle I_{T}^{-}\right\rangle $
and $\left\langle I_{2}\right\rangle =\left\langle I_{T}^{+}\right\rangle $. 

Not so in a finite geometry: there are of course reflections at the
boundaries.

It is possible to relate the two sets of operators $I_{1}$, $I_{2}$
and $I_{T}^{-}$, $I_{T}^{+}$. The special case of quantized resistances
$Z_{S}=Z_{D}=\frac{R_{0}}{2}=\frac{h}{2e^{2}}$ was previously discussed
by the author and collaborators in \cite{p1}.

Using their definition one has in particular that the total tunneling
current is: \[
I_{total}=I_{1}+I_{2}=I_{T}^{+}+I_{T}^{-}\]
We now use the boundary conditions, eq.(\ref{bc}) : \begin{align*}
I_{1} & =\frac{1}{2eZ_{S}}\left(\mu_{L}^{+}+\mu_{L}^{-}\right)=\frac{Z_{0}}{Z_{S}}\left(i_{L}^{+}+i_{L}^{-}\right)\\
I_{2} & =\frac{1}{2eZ_{D}}\left(\mu_{R}^{+}+\mu_{R}^{-}\right)=\frac{Z_{0}}{Z_{D}}\left(i_{R}^{+}+i_{R}^{-}\right)\end{align*}
 where the source and drain voltages have been set to the ground in
this geometry; we have also used the fact that the operators are uniform. 

This implies: \[
Z_{S}I_{1}-Z_{D}I_{2}=Z_{0}\left(I_{T}^{-}-I_{T}^{+}\right)\]
 Finally: \begin{align}
\left(\begin{array}{cc}
Z_{S} & -Z_{D}\\
1 & 1\end{array}\right)\left(\begin{array}{c}
I_{1}\\
I_{2}\end{array}\right) & =\left(\begin{array}{cc}
Z_{0} & -Z_{0}\\
1 & 1\end{array}\right)\left(\begin{array}{c}
I_{-T}\\
I_{T}^{+}\end{array}\right)\nonumber \\
\left(\begin{array}{c}
I_{1}\\
I_{2}\end{array}\right) & =\frac{1}{Z_{S}+Z_{D}}\left(\begin{array}{cc}
Z_{D}+Z_{0} & Z_{D}-Z_{0}\\
Z_{S}-Z_{0} & Z_{S}+Z_{0}\end{array}\right)\left(\begin{array}{c}
I_{T}^{-}\\
I_{T}^{+}\end{array}\right)\label{cur-ren}\end{align}

We stress that these relations hold very generally: the tunnel currents
need not be small; the range of validity extends from the weak tunneling
to the strong tunneling regime. 

Essential remarks: 

- (i) We recover the intuitively expected result that IF impedance
matching is realized at BOTH boundaries, namely: \[
Z_{S}=Z_{0}=Z_{D}\]
 Then the measured currents and the injected currents are identical:
\[
\left(\begin{array}{c}
I_{1,matched}\\
I_{2,matched}\end{array}\right)=\left(\begin{array}{c}
I_{T}^{-}\\
I_{T}^{+}\end{array}\right).\]
 This gives the physical meaning of the injected current operators
$I_{T}^{-}$ and $I_{T}^{+}$: they would be the current operators
if the system were infinite. This means that we can view the relation
between the operators for the injected currents $I_{T}^{-}$ and $I_{T}^{+}$and
the operators for the measured currents $I_{1}$, $I_{2}$ as an operator
renormalization when one goes from the infinite system to the finite-length
system.

- (ii) Why is it useful to consider the operators $I_{T}^{-}$ and
$I_{T}^{+}$? Because in the DC regime by making some reasonable assumptions
(Poisson injection by the STM tip) the Fano ratios are quite easy
to find: actually they are identical with those of the infinite system,
which is not quite unreasonable given the fact that $I_{T}^{-}$ and
$I_{T}^{+}$ are just the current operators of the impedance-matched
system. 

For AC excess noise the correlators for $I_{T}^{-}$ and $I_{T}^{+}$
are not so easily found. But still we will find that assuming that
their correlators are unchanged from their values in the infinite
system yields the dominant behaviour for the excess noise for the
measured currents $I_{1}$, $I_{2}$.

\subsubsection{Physical interpretation : reflections.}

The renormalization from $I_{T}^{-}$ and $I_{T}^{+}$to $I_{1}$
and $I_{2}$ physically results from multiple reflections back and
forth at the boundaries. This is a completely classical effect as
seen for instance in waveguides, sound waves in a tube, etc, whenever
load impedances are connected to the boundaries of the system and
whenever there is an impedance mismatch. 

Since this relation is the building-block of this paper for the sake
of pedagogy we will now show how it can be recovered by directly considering
reflections of the plasma wave. Let us define reflections coefficients
$r_{S}$ and $r_{D}$. 

The plasma wave has two chiral components on the left and on the right
of the impurity $i_{R/L}^{\pm}$. We will build these currents sequentially. 

\begin{figure}
\includegraphics[%
  scale=0.8]{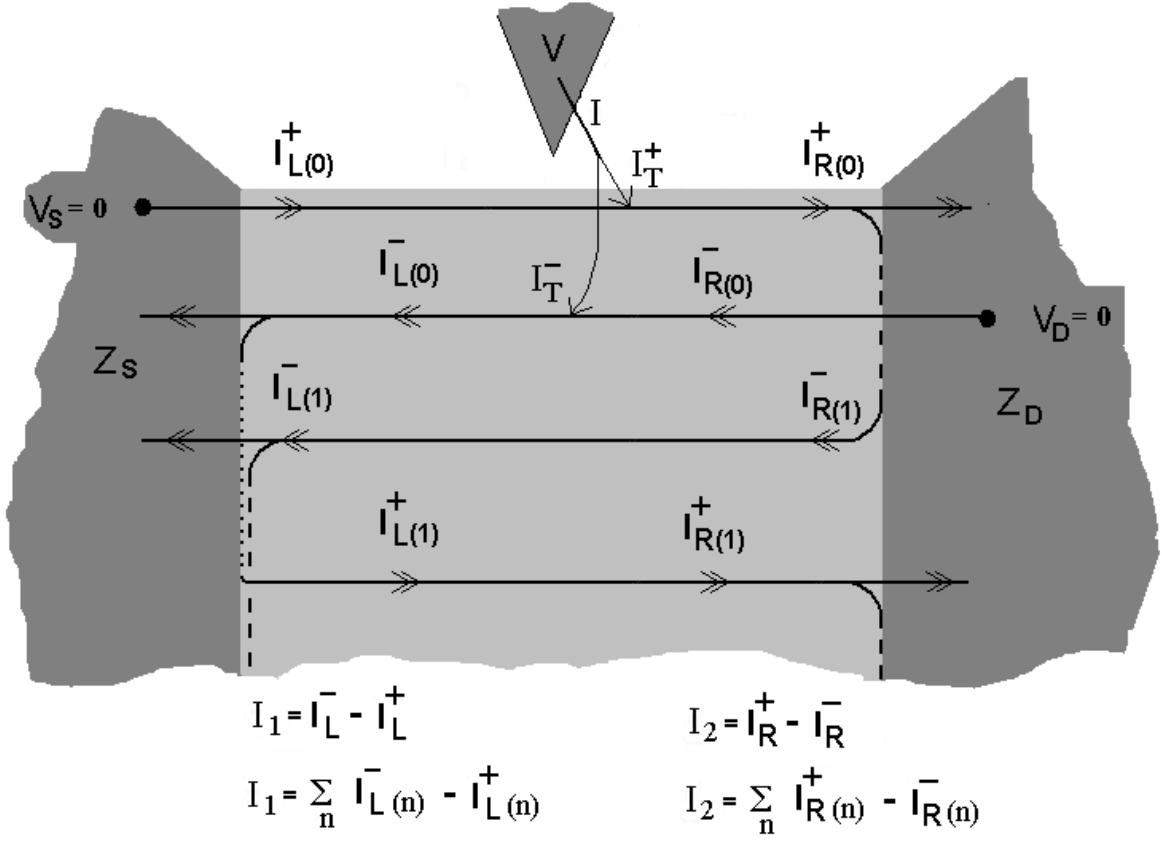}

\caption{\label{cap:reflections-stm}Reflections at the boundaries for the
currents injected by a STM tip.}
\end{figure}

\textbf{Order zero:}

If we take no account of the reflections then at zero order in a development
in the reflection coefficients: \begin{align*}
i_{L(0)}^{+} & =0\qquad i_{R(0)}^{-}=0\\
i_{R(0)}^{+} & =I_{T}^{+}\qquad i_{L(0)}^{-}=I_{T}^{-}\end{align*}

\textbf{Order one:}\begin{align*}
i_{L(1)}^{+} & =r_{D}~i_{L(0)}^{-}\qquad i_{R(1)}^{-}=r_{S}~i_{R(0)}^{+}\\
i_{R(1)}^{+} & =i_{L(1)}^{+}\qquad i_{L(1)}^{-}=i_{R(1)}^{-}\end{align*}
 The second line follows just from current conservation. 

\textbf{Order} $n$: \begin{align*}
i_{L(n)}^{+} & =r_{D}~i_{L(n-1)}^{-}\qquad i_{R(n)}^{-}=r_{S}~i_{R(n-1)}^{+}\\
i_{R(n)}^{+} & =i_{L(n)}^{+}\qquad i_{L(n)}^{-}=i_{R(n)}^{-}\end{align*}
 Therefore for $n\geq1$: \[
\left(\begin{array}{c}
i_{R(n)}^{+}\\
i_{L(n)}^{-}\end{array}\right)=\left(\begin{array}{cc}
0 & r_{S}\\
r_{D} & 0\end{array}\right)\left(\begin{array}{c}
i_{R(n-1)}^{+}\\
i_{L(n-1)}^{-}\end{array}\right)\]
 and for $n\geq2$: \[
\left(\begin{array}{c}
i_{L(n)}^{+}\\
i_{R(n)}^{-}\end{array}\right)=\left(\begin{array}{cc}
0 & r_{S}\\
r_{D} & 0\end{array}\right)\left(\begin{array}{c}
i_{L(n-1)}^{+}\\
i_{R(n-1)}^{-}\end{array}\right).\]
 Defining $M=\left(\begin{array}{cc}
0 & r_{S}\\
r_{D} & 0\end{array}\right)$ we have: later we will physically interpret some of this paper's
results for the shot noise through similar heuristic reasonings\begin{align*}
\left(\begin{array}{c}
i_{R}^{+}\\
i_{L}^{-}\end{array}\right) & =\left[1+M+M^{2}+M^{3}+...\right]\left(\begin{array}{c}
i_{R(0)}^{+}\\
i_{L(0)}^{-}\end{array}\right)\\
 & =\left[1-M\right]^{-1}\left(\begin{array}{c}
I_{+T}\\
I_{-T}\end{array}\right)\\
 & =\frac{1}{1-r_{S}r_{D}}\left(\begin{array}{cc}
1 & r_{S}\\
r_{D} & 1\end{array}\right)\left(\begin{array}{c}
I_{T}^{+}\\
I_{T}^{-}\end{array}\right)\end{align*}
 and \begin{align*}
\left(\begin{array}{c}
i_{L}^{+}\\
i_{R}^{-}\end{array}\right) & =\left[1+M+M^{2}+M^{3}+...\right]\left(\begin{array}{c}
i_{L(1)}^{+}\\
i_{R(1)}^{-}\end{array}\right)\\
 & =\left[1-M\right]^{-1}\left(\begin{array}{c}
r_{S}~I_{T}^{-}\\
r_{D}~I_{T}^{+}\end{array}\right)\end{align*}
 Since $I_{1}=i_{L}^{-}-i_{L}^{+}$ and $I_{2}=i_{R}^{+}-i_{R}^{-}$
straightforward calculations lead to: \[
\left(\begin{array}{c}
I_{1}\\
I_{2}\end{array}\right)=\frac{1}{1-r_{S}r_{D}}\left(\begin{array}{cc}
1-r_{S} & \left(1-r_{S}\right)r_{D}\\
\left(1-r_{D}\right)r_{S} & 1-r_{D}\end{array}\right)\left(\begin{array}{c}
I_{T}^{-}\\
I_{T}^{+}\end{array}\right)\]
 Comparison with the results found above: \[
\left(\begin{array}{c}
I_{1}\\
I_{2}\end{array}\right)=\frac{1}{Z_{S}+Z_{D}}\left(\begin{array}{cc}
Z_{D}+Z_{0} & Z_{D}-Z_{0}\\
Z_{S}-Z_{0} & Z_{S}+Z_{0}\end{array}\right)\left(\begin{array}{c}
I_{T}^{-}\\
I_{T}^{+}\end{array}\right)\]
 shows they are identical provided one identifies \[
r_{D}=\frac{Z_{D}-Z_{0}}{Z_{D}+Z_{0}},\qquad r_{S}=\frac{Z_{S}-Z_{0}}{Z_{S}+Z_{0}}\]
 \emph{which is just the expression expected for e.g. a sound wave
in a tube terminated by two load impedances}! 

This shows that the renormalizations of the tunneling currents follow
simply from multiple reflections at the boundaries of the Luttinger
liquid. Observe again that the origin of the phenomenon is \textit{purely
classical} and has no quantum grounds.

\subsubsection{Excess noise for $I_{T}^{-}$ and $I_{T}^{+}$ for an STM with asymetric
injection\label{sub:Excess-noise-for-injected}.}

There are two processes for the injection: either (i) injection of
an electron at $k_{F}$ or (ii) injection at $-k_{F}$. 

As shown in \cite{fractionalization} these electrons split in the
LL and fractional eigenstates are created so that physically one injects
the charges\begin{align*}
Q_{+} & =\frac{1+K}{2}\\
Q_{-} & =\frac{1-K}{2}\end{align*}
 in each arm: process (i) injection at $k_{F}$ : $Q_{+}$ to the
right and $Q_{-}$ to the left; process (ii): $Q_{-}$ to the right
and $Q_{+}$ to the left. 

The total number of injections by either processes obeys Poisson statistics:
we will work in a weak transmission limit (the current injected goes
as a power law of the voltage difference between the electrode tip
and the nanotube). We define the probabilities that a given injection
is by process (ii) rather than (i) by $T$. Microscopically we write
the usual Luttinger hamiltonian plus a tunneling hamiltonian: \[
V=\left(\Gamma_{+k_{F}}\Psi_{R}+\Gamma_{-k_{F}}\Psi_{L}~\right)c^{+}+h.c.\]
 $\Psi_{R}$ is the electron operator for an electron at the right
Fermi point $+k_{F}$ and $\Psi_{L}$ is the electron operator for
an electron at the left Fermi point $+k_{F}$; $c^{+}$ is the creation
operator for an electron in the electrode. We have allowed for distinct
probability amplitudes for the injection of left and right electrons
for full generality. The probabilities $T$ and $R$ are then simply:
\begin{align*}
R & =\frac{\left|\Gamma_{+k_{F}}\right|^{2}}{\left|\Gamma_{+k_{F}}\right|^{2}+\left|\Gamma_{-k_{F}}\right|^{2}}\\
T & =\frac{\left|\Gamma_{-k_{F}}\right|^{2}}{\left|\Gamma_{+k_{F}}\right|^{2}+\left|\Gamma_{-k_{F}}\right|^{2}}\end{align*}
 $T$ is therefore the probability that a given charge injection is
done with a left Fermi point electron rather than with a right Fermi
electron. 

As already emphasized by Cr\'{e}pieux et al, \cite{stm} injection
in a LL through an STM tip works as a Hanbury-Brown and Twiss device.
Some care is however needed in the comparison: in a standard Hanbury-Brown
and Twiss setting a source signal is partitioned with a probability
$T$ to be transmitted and a probability $R=1-T$ of reflection; here
for the LL what plays the role of the partitioning is the fact that
an electron can be injected at either $+k_{F}$ or $-k_{F}$: \emph{it
is not the splitting of charge into fractional charges which acts
as a partitioning}. Choosing an asymetric STM electrode allows a better
comparison to the Hanbury-Brown and Twiss setting since the probabilities
$T$ and $R$ are not fixed at the value $T=R=1/2$ as in a symetric
electrode. In the present experimental state of the art it may sem
far-fetched to realize selective or asymetric injection of electrons
but it might be feasible in a foreseeable future in quantum wires. 

If the total number of injections is $N$: \[
\left\langle \Delta N^{2}\right\rangle -\left\langle N\right\rangle =0\]
 since we have assumed Poissonian statistics for the total number
of electron injections. We call $m$ and $m^{\prime}$ the number
of injections by respectively process (ii) or (i) above. Evidently
one has the partition noise result (Burgess variance theorem): \begin{align*}
\left\langle m^{2}\right\rangle -\left\langle m\right\rangle ^{2} & =T^{2}\left\langle \Delta N^{2}\right\rangle +T(1-T)\left\langle N\right\rangle \\
\left\langle \Delta m\Delta m^{\prime}\right\rangle  & =T(1-T)\left(\left\langle \Delta N^{2}\right\rangle -\left\langle N\right\rangle \right).\end{align*}
 Now using the assumption that the total number of injections obeys
Poisson statistics, one finds that:\begin{align*}
\left\langle \Delta m^{2}\right\rangle  & =T\left\langle N\right\rangle =\left\langle m\right\rangle \\
\left\langle \Delta m^{\prime2}\right\rangle  & =R\left\langle N\right\rangle =\left\langle m^{\prime}\right\rangle \\
\left\langle \Delta m\Delta m^{\prime}\right\rangle  & =0.\end{align*}
 We infer the charge $N_{T}^{-}=Q_{+}m+Q_{-}m^{\prime}$ transmitted
to the left of the tip (arm $1$), the current injected in the left
arm $I_{T}^{-}$ and its fluctuations: \begin{align}
\left\langle N_{T}^{-}\right\rangle  & =Q_{+}\left\langle m\right\rangle +Q_{-}\left\langle m^{\prime}\right\rangle =(Q_{+}T+Q_{-}R)\left\langle N\right\rangle \nonumber \\
\left\langle I_{T}^{-}\right\rangle  & =(Q_{+}T+Q_{-}R)\left\langle I\right\rangle \label{i-t}\\
\left\langle \left(\Delta N_{T}^{-}\right)^{2}\right\rangle  & =\left\langle \left(Q_{+}\Delta m+Q_{-}\Delta m^{\prime}\right)^{2}\right\rangle \nonumber \\
 & =Q_{+}^{2}\left\langle \Delta m^{2}\right\rangle +Q_{-}^{2}\left\langle \Delta m^{\prime2}\right\rangle =(Q_{+}^{2}T+Q_{-}^{2}R)\left\langle N\right\rangle \nonumber \\
F_{1} & =\frac{\left\langle \left(\Delta I_{T}^{-}\right)^{2}\right\rangle }{\left\langle I_{T}^{-}\right\rangle }=\frac{\left\langle \left(\Delta N_{T}^{-}\right)^{2}\right\rangle }{\left\langle N_{T}^{-}\right\rangle }=\frac{Q_{+}^{2}T+Q_{-}^{2}R}{Q_{+}T+Q_{-}R}\label{f1}\end{align}
 And likewise in the right arm (arm $2$): \begin{align}
\left\langle N_{T}^{+}\right\rangle  & =Q_{-}\left\langle m\right\rangle +Q_{+}\left\langle m^{\prime}\right\rangle =(Q_{-}T+Q_{+}R)\left\langle N\right\rangle \nonumber \\
\left\langle I_{T}^{+}\right\rangle  & =(Q_{-}T+Q_{+}R)\left\langle I\right\rangle \label{i+t}\\
\left\langle \left(\Delta N_{T}^{+}\right)^{2}\right\rangle  & =\left\langle \left(Q_{-}\Delta m+Q_{+}\Delta m^{\prime}\right)^{2}\right\rangle \nonumber \\
 & =Q_{-}^{2}\left\langle \Delta m^{2}\right\rangle +Q_{+}^{2}\left\langle \Delta m^{\prime2}\right\rangle =(Q_{-}^{2}T+Q_{+}^{2}R)\left\langle N\right\rangle \nonumber \\
F_{2} & =\frac{\left\langle \left(\Delta I_{T}^{+}\right)^{2}\right\rangle }{\left\langle I_{T}^{+}\right\rangle }=\frac{\left\langle \left(\Delta N_{T}^{+}\right)^{2}\right\rangle }{\left\langle N_{T}^{+}\right\rangle }=\frac{Q_{-}^{2}T+Q_{+}^{2}R}{Q_{-}T+Q_{+}R}\label{f2}\end{align}
 The cross correlations follow easily: \begin{align}
\left\langle \Delta N_{T}^{-}\Delta N_{T}^{+}\right\rangle  & =\left\langle \left(Q_{+}\Delta m+Q_{-}\Delta m^{\prime}\right)\left(Q_{-}\Delta m+Q_{+}\Delta m^{\prime}\right)\right\rangle \nonumber \\
 & =Q_{+}Q_{-}\left(\left\langle \Delta m^{2}\right\rangle +\left\langle \Delta m^{\prime2}\right\rangle \right)\nonumber \\
 & =Q_{+}Q_{-}\left\langle N\right\rangle \nonumber \\
\left\langle \Delta I_{T}^{-}\Delta I_{T}^{+}\right\rangle  & =Q_{+}Q_{-}\left\langle I_{T}^{-}+I_{T}^{+}\right\rangle \nonumber \\
\frac{\left\langle \Delta I_{T}^{-}\Delta I_{T}^{+}\right\rangle }{\left\langle I_{T}^{-}\right\rangle } & =\frac{Q_{+}Q_{-}}{Q_{+}T+Q_{-}R},~\nonumber \\
\frac{\left\langle \Delta I_{T}^{-}\Delta I_{T}^{+}\right\rangle }{\left\langle I_{T}^{+}\right\rangle } & =\frac{Q_{+}Q_{-}}{Q_{-}T+Q_{+}R},~\nonumber \\
\frac{\left\langle \Delta I_{T}^{-}\Delta I_{T}^{+}\right\rangle }{\sqrt{\left\langle I_{T}^{-}\right\rangle \left\langle I_{T}^{+}\right\rangle }} & =\frac{Q_{+}Q_{-}}{\sqrt{\left(Q_{+}T+Q_{-}R\right)\left(Q_{-}T+Q_{+}R\right)}}\label{f12}\end{align}

We observe in passing that they are \emph{positive} a fact already
pointed out in \cite{stm} for the special case of symetric injection.
All those straightforward results can of course be recovered by a
lengthier route through the Keldysh technique following Cr\'{e}pieux
et al. 

We stress furthermore that these relations are valid whether the LL
wire is finite or not. Although the exact values of the correlators
and of the current averages depend on the length of the system (the
quantum average being taken over the length-dependent ground state)
the Fano ratios are clearly invariant. In particular this means that
the Fano ratios for the injected currents are \emph{exactly} those
of the measured currents $I_{1}$ and $I_{2}$ in the infinite system.

\subsubsection{Excess noise for the measured currents: a mixing of direct and cross-correlations
as compared to the infinite system\label{sub:Excess-noise-for-measured}. }

Since : \[
\left(\begin{array}{c}
I_{1}\\
I_{2}\end{array}\right)=\frac{1}{Z_{S}+Z_{D}}\left(\begin{array}{cc}
Z_{D}+Z_{0} & Z_{D}-Z_{0}\\
Z_{S}-Z_{0} & Z_{S}+Z_{0}\end{array}\right)\left(\begin{array}{c}
I_{T}^{-}\\
I_{T}^{+}\end{array}\right)\]
 it follows that the DC noise correlations for the measured currents
are: \begin{align}
\left\langle \left(\Delta I_{1}\right)^{2}\right\rangle  & =\frac{1}{\left(Z_{S}+Z_{D}\right)^{2}}\left[\begin{array}{c}
\left(Z_{D}+Z_{0}\right)^{2}\left\langle \left(\Delta I_{T}^{-}\right)^{2}\right\rangle +\left(Z_{D}-Z_{0}\right)^{2}\left\langle \left(\Delta I_{T}^{+}\right)^{2}\right\rangle +2\left(Z_{D}^{2}-Z_{0}^{2}\right)\left\langle \Delta I_{T}^{+}~\Delta I_{T}^{-}\right\rangle \end{array}\right]\label{cross}\\
\left\langle \left(\Delta I_{2}\right)^{2}\right\rangle  & =\frac{1}{\left(Z_{S}+Z_{D}\right)^{2}}\left[\begin{array}{c}
\left(Z_{S}-Z_{0}\right)^{2}\left\langle \left(\Delta I_{T}^{-}\right)^{2}\right\rangle \end{array}+\left(Z_{S}+Z_{0}\right)^{2}\left\langle \left(\Delta I_{T}^{+}\right)^{2}\right\rangle +2\left(Z_{S}^{2}-Z_{0}^{2}\right)\left\langle \Delta I_{T}^{+}~\Delta I_{T}^{-}\right\rangle \right]\\
\left\langle \Delta I_{1}~\Delta I_{2}\right\rangle  & =\frac{\left(Z_{D}+Z_{0}\right)\left(Z_{S}-Z_{0}\right)\left\langle \left(\Delta I_{T}^{-}\right)^{2}\right\rangle +\left(Z_{S}+Z_{0}\right)\left(Z_{D}-Z_{0}\right)\left\langle \left(\Delta I_{T}^{+}\right)^{2}\right\rangle +2\left(Z_{S}Z_{D}+Z_{0}^{2}\right)\left\langle \Delta I_{T}^{+}~\Delta I_{T}^{-}\right\rangle }{\left(Z_{S}+Z_{D}\right)^{2}}\end{align}
 where $\left\langle \Delta I_{T}^{\pm}~\Delta I_{T}^{\pm}\right\rangle $
are the correlators of the injected currents. 

Since the Fano factors for$\left\langle \Delta I_{T}^{\pm}~\Delta I_{T}^{\pm}\right\rangle $
are identical to those of $I_{1}$ and $I_{2}$ in the infinite system
this shows that in the presence of boundaries there is a mixing of
what would be in the infinite system the direct and cross correlations. 

Since according to eq.(\ref{i+t},\ref{i-t}): 

\begin{align}
\left\langle I_{T}^{-}\right\rangle  & =(Q_{+}T+Q_{-}R)\left\langle I\right\rangle ,\nonumber \\
\left\langle I_{T}^{+}\right\rangle  & =(Q_{-}T+Q_{+}R)\left\langle I\right\rangle .\label{tunnel}\end{align}
 The latter equations then imply that (using eq.(\ref{cur-ren}) ): 

\begin{align}
\left(\begin{array}{c}
I_{1}\\
I_{2}\end{array}\right) & =\frac{I}{Z_{S}+Z_{D}}\left(\begin{array}{c}
Z_{D}+Z_{0}\left(Q_{+}-Q_{-}\right)\left(T-R\right)\\
Z_{S}+Z_{0}\left(Q_{+}-Q_{-}\right)\left(R-T\right)\end{array}\right)\label{i}\\
 & =\frac{I_{T}^{-}}{(Q_{+}T+Q_{-}R)\left(Z_{S}+Z_{D}\right)}\left(\begin{array}{c}
Z_{D}+Z_{0}\left(Q_{+}-Q_{-}\right)\left(T-R\right)\\
Z_{S}+Z_{0}\left(Q_{+}-Q_{-}\right)\left(R-T\right)\end{array}\right)\end{align}
 Gathering eq.(\ref{cur-ren}) and eq.(\ref{cross}) one then finds
immediately:\begin{align*}
\left\langle \left(\Delta I_{1}\right)^{2}\right\rangle  & =\frac{\left(Z_{D}+Z_{0}\right)^{2}\frac{Q_{+}^{2}T+Q_{-}^{2}R}{Q_{+}T+Q_{-}R}+\left(Z_{D}-Z_{0}\right)^{2}\frac{Q_{-}^{2}T+Q_{+}^{2}R}{Q_{+}T+Q_{-}R}+2\left(Z_{D}^{2}-Z_{0}^{2}\right)\frac{Q_{+}Q_{-}}{Q_{+}T+Q_{-}R}}{\left(Z_{S}+Z_{D}\right)^{2}}\left\langle I_{T}^{-}\right\rangle \\
\left\langle \left(\Delta I_{2}\right)^{2}\right\rangle  & =\frac{\left(Z_{S}-Z_{0}\right)^{2}\frac{Q_{+}^{2}T+Q_{-}^{2}R}{Q_{+}T+Q_{-}R}+\left(Z_{S}+Z_{0}\right)^{2}\frac{Q_{-}^{2}T+Q_{+}^{2}R}{Q_{+}T+Q_{-}R}+2\left(Z_{S}^{2}-Z_{0}^{2}\right)\frac{Q_{+}Q_{-}}{Q_{+}T+Q_{-}R}}{\left(Z_{S}+Z_{D}\right)^{2}}\left\langle I_{T}^{-}\right\rangle \\
\left\langle \Delta I_{1}~\Delta I_{2}\right\rangle  & =\frac{\left(Z_{D}+Z_{0}\right)\left(Z_{S}-Z_{0}\right)\frac{Q_{+}^{2}T+Q_{-}^{2}R}{Q_{+}T+Q_{-}R}+\left(Z_{S}+Z_{0}\right)\left(Z_{D}-Z_{0}\right)\frac{Q_{-}^{2}T+Q_{+}^{2}R}{Q_{+}T+Q_{-}R}+2\left(Z_{S}Z_{D}+Z_{0}^{2}\right)\frac{Q_{+}Q_{-}}{Q_{+}T+Q_{-}R}}{\left(Z_{S}+Z_{D}\right)^{2}}\left\langle I_{T}^{-}\right\rangle \end{align*}
 and plugging the expression of the current injected to the left in
function of the total current from eq.(\ref{tunnel}): \begin{align*}
\left\langle \left(\Delta I_{1}\right)^{2}\right\rangle  & =\frac{1}{\left(Z_{S}+Z_{D}\right)^{2}}\left[\begin{array}{c}
\left(Z_{D}+Z_{0}\right)^{2}\left(Q_{+}^{2}T+Q_{-}^{2}R\right)\\
+\left(Z_{D}-Z_{0}\right)^{2}\left(Q_{-}^{2}T+Q_{+}^{2}R\right)+2\left(Z_{D}^{2}-Z_{0}^{2}\right)\left(Q_{+}Q_{-}\right)\end{array}\right]\left\langle I\right\rangle \\
\left\langle \left(\Delta I_{2}\right)^{2}\right\rangle  & =\frac{1}{\left(Z_{S}+Z_{D}\right)^{2}}\left[\begin{array}{c}
\left(Z_{S}-Z_{0}\right)^{2}\left(Q_{+}^{2}T+Q_{-}^{2}R\right)+\\
\left(Z_{S}+Z_{0}\right)^{2}\left(Q_{-}^{2}T+Q_{+}^{2}R\right)+2\left(Z_{S}^{2}-Z_{0}^{2}\right)\left(Q_{+}Q_{-}\right)\end{array}\right]\left\langle I\right\rangle \\
\left\langle \Delta I_{1}~\Delta I_{2}\right\rangle  & =\frac{1}{\left(Z_{S}+Z_{D}\right)^{2}}\left[\begin{array}{c}
\left(Z_{D}+Z_{0}\right)\left(Z_{S}-Z_{0}\right)\left(Q_{+}^{2}T+Q_{-}^{2}R\right)\\
+\left(Z_{S}+Z_{0}\right)\left(Z_{D}-Z_{0}\right)\left(Q_{-}^{2}T+Q_{+}^{2}R\right)+2\left(Z_{S}Z_{D}+Z_{0}^{2}\right)\left(Q_{+}Q_{-}\right)\end{array}\right]\left\langle I\right\rangle \end{align*}
 or in terms of the currents in each branch (using eq.(\ref{i}) above): 

\begin{align}
\left\langle \left(\Delta I_{1}\right)^{2}\right\rangle  & =\frac{\left(Z_{D}+Z_{0}\right)^{2}\left(Q_{+}^{2}T+Q_{-}^{2}R\right)+\left(Z_{D}-Z_{0}\right)^{2}\left(Q_{-}^{2}T+Q_{+}^{2}R\right)+2\left(Z_{D}^{2}-Z_{0}^{2}\right)\left(Q_{+}Q_{-}\right)}{\left(Z_{S}+Z_{D}\right)}\frac{\left\langle I_{1}\right\rangle }{Z_{D}+Z_{0}\left(Q_{+}-Q_{-}\right)\left(T-R\right)}\nonumber \\
\left\langle \left(\Delta I_{2}\right)^{2}\right\rangle  & =\frac{\left(Z_{S}-Z_{0}\right)^{2}\left(Q_{+}^{2}T+Q_{-}^{2}R\right)+\left(Z_{S}+Z_{0}\right)^{2}\left(Q_{-}^{2}T+Q_{+}^{2}R\right)+2\left(Z_{S}^{2}-Z_{0}^{2}\right)\left(Q_{+}Q_{-}\right)}{\left(Z_{S}+Z_{D}\right)}\frac{\left\langle I_{2}\right\rangle }{Z_{S}+Z_{0}\left(Q_{+}-Q_{-}\right)\left(R-T\right)}\nonumber \\
\left\langle \Delta I_{1}~\Delta I_{2}\right\rangle  & =\frac{\left(Z_{D}+Z_{0}\right)\left(Z_{S}-Z_{0}\right)\left(Q_{+}^{2}T+Q_{-}^{2}R\right)+\left(Z_{S}+Z_{0}\right)\left(Z_{D}-Z_{0}\right)\left(Q_{-}^{2}T+Q_{+}^{2}R\right)+2\left(Z_{S}Z_{D}+Z_{0}^{2}\right)\left(Q_{+}Q_{-}\right)}{\left(Z_{S}+Z_{D}\right)^{2}}\left\langle I\right\rangle \label{result}\end{align}
 which is the main result of the section. 

Up to now we stress that the exact values of the charges $Q_{+}$
and $Q_{-}$ of the fractional states have not been used in the calculations:
this implies that the set of equations eq.(\ref{result}) can be used
to extract experimentally their values EVEN in the absence of impedance
matching ONCE the values of the load impedances $Z_{S}$ and $Z_{D}$
and the characteristic impedance $Z_{0}$ are known through any transport
measurement (time domain reflectometry, DC or AC conductance, etc...).
Since one has three correlators, the experimental measurement of two
of them should in principle allow for an extraction of the two charges
$Q_{+}$ and $Q_{-}$ by fitting their values \emph{while} the measurement
of the third correlator (e.g. the cross-correlation) then becomes
a distinct non-trivial prediction of the theory WITHOUT any fit. 

However such a straightforward approach faces us with one conceptual
issue: if we use the distinct predictions of the LL theory \cite{fractionalization,p3}
the characteristic impedance $Z_{0}$ and the charges $Q_{+}$ and
$Q_{-}$ are found not to be independent since:\[
Q_{+}=\frac{1+K}{2},~Q_{-}=\frac{1-K}{2},~Z_{0}=\frac{h}{2Ke^{2}}.\]
 If we plug in these values in the expression for the current correlators
one gets: \begin{align}
\left\langle \left(\Delta I_{1}\right)^{2}\right\rangle  & =\frac{Z_{D}^{2}+\frac{R_{0}^{2}}{4}+Z_{D}R_{0}\left(T-R\right)}{\left(Z_{S}+Z_{D}\right)\left(Z_{D}+\frac{R_{0}}{2}\left(T-R\right)\right)}\left\langle I_{1}\right\rangle \nonumber \\
\left\langle \left(\Delta I_{2}\right)^{2}\right\rangle  & =\frac{Z_{S}^{2}+\frac{R_{0}^{2}}{4}-Z_{S}R_{0}\left(T-R\right)}{\left(Z_{S}+Z_{D}\right)\left(Z_{S}-\frac{R_{0}}{2}\left(T-R\right)\right)}\left\langle I_{2}\right\rangle \nonumber \\
\left\langle \Delta I_{1}~\Delta I_{2}\right\rangle  & =\frac{Z_{S}Z_{D}-\frac{R_{0}^{2}}{4}+(Z_{S}-Z_{D})\frac{R_{0}}{2}\left(T-R\right)}{\left(Z_{S}+Z_{D}\right)^{2}}\left\langle I\right\rangle \label{fortuitous}\end{align}
 where $R_{0}=h/e^{2}$ is the quantum of resistance. All the dependence
on the Luttinger parameter $K$ has vanished: equations Eq.(\ref{fortuitous})
are therefore correct even for free fermions ($K=1$) provided they
are connected to two load impedances $Z_{S}$ and $Z_{D}$ \textit{and
that all phase coherence effects are neglected}. It might seem therefore
that the strong interaction physics can not be probed in this manner. 

The fractional charges \emph{seemingly} (there is one proviso) can
not be measured in a DC experiment: this generalizes the conclusion
reached by \cite{stm}, whose calculations we recover as a subcase
of ours with a symetric setup ($T=R=1/2$) for the special choice
of:\[
Z_{S}=Z_{D}=\frac{R_{0}}{2}=\frac{h}{2e^{2}}\]
 as: \begin{align}
\left\langle \left(\Delta I_{1}\right)^{2}\right\rangle  & =\left\langle \Delta I_{1}\right\rangle \\
\left\langle \left(\Delta I_{2}\right)^{2}\right\rangle  & =\left\langle \Delta I_{2}\right\rangle \\
\left\langle \Delta I_{1}~\Delta I_{2}\right\rangle  & =0\end{align}
 However we can see that the fact that these correlators take the
same value as in an infinite non-interacting system results actually
from the conspiracy of three elements: (1) the fortuitous cancellation
of the characteristic impedance with the fractional charges; (2) the
fact that the inhomogeneous LL (LL with Fermi leads) and related models
imply that the load impedances take a not innocuous value: $Z_{S}=Z_{D}=\frac{R_{0}}{2}=\frac{h}{2e^{2}}$;
(3) the symetric injection of $+k_{F}$ and $-k_{F}$ electrons resulting
in: $T=R=1/2$. 

The \emph{proviso} to the negative conclusion reached here is to use
impedance matching: since reflections are killed one is sure to work
with an effectively infinite Luttinger liquid and there can then be
no ambiguity on the interpretation of shot noise experiments. Indeed
for the matched system the identity of the measured currents with
the currents injected ensures that we are measuring intrinsic properties
of the LL unspoiled by reflections; namely:\[
\left(\begin{array}{c}
I_{1,matched}\\
I_{2,matched}\end{array}\right)=\left(\begin{array}{c}
I_{T}^{-}\\
I_{T}^{+}\end{array}\right)\]
 implies immediately (see section \ref{sub:Excess-noise-for-injected})
that the current correlators of the matched system coincide with those
of the infinite system: \begin{align}
\frac{\left\langle \Delta I_{1,matched}^{2}\right\rangle }{\left\langle \Delta I_{1,matched}\right\rangle } & =\frac{Q_{+}^{2}T+Q_{-}^{2}R}{Q_{+}T+Q_{-}R},\\
\frac{\left\langle \Delta I_{2,matched}^{2}\right\rangle }{\left\langle \Delta I_{2,matched}\right\rangle } & =\frac{Q_{-}^{2}T+Q_{+}^{2}R}{Q_{-}T+Q_{+}R},\\
\frac{\left\langle \Delta I_{1,matched}\Delta I_{2,matched}\right\rangle }{\left\langle \Delta I_{1,matched}\right\rangle } & =\frac{Q_{+}Q_{-}}{Q_{+}T+Q_{-}R}.\end{align}

The final conclusion of this section is therefore that in general
the effective charges of the fractional states created upon injection
of charge by a tunneling STM tip are unobservable in a DC experiment
UNLESS impedance matching is realized.

\subsection{AC shot noise.}

\subsubsection{Renormalization of the injected currents into measured currents.}

The previous relations for the renormalization of the injected currents
are only valid in the DC regime (e.g. for the zero frequency Fourier
components of the current): we want to derive a similar relation for
the non-zero Fourier components, since this will enable us to discuss
AC shot noise. We call the length to the right and to the left of
the tunneling point $L_{2}$ and $L_{1}$. We work at frequency $\omega$
and define the plasma wave phases:

\[
\phi_{2}=\frac{\omega}{u}L_{2}\qquad\phi_{1}=\frac{\omega}{u}L_{1}.\]
 The previous result is then modified as: \begin{align}
\left(\begin{array}{c}
I_{1}(\omega)\\
I_{2}(\omega)\end{array}\right) & =\frac{2Z_{0}}{\left(Z_{S}+Z_{0}\right)\left(Z_{D}+Z_{0}\right)\exp-i\left(\phi_{1}+\phi_{2}\right)-\left(Z_{S}-Z_{0}\right)\left(Z_{D}-Z_{0}\right)\exp i\left(\phi_{1}+\phi_{2}\right)}\label{ac-ren}\\
 & \times\left(\begin{array}{cc}
\left(Z_{D}+Z_{0}\right)e^{-i\left(\phi_{1}+\phi_{2}\right)} & \left(Z_{D}-Z_{0}\right)\\
\left(Z_{S}-Z_{0}\right) & \left(Z_{S}+Z_{0}\right)e^{-i\left(\phi_{1}+\phi_{2}\right)}\end{array}\right)\left(\begin{array}{c}
I_{T}^{-}(-L_{1},\omega)\\
I_{T}^{+}(L_{2},\omega)\end{array}\right)\nonumber \end{align}
 where $\left(\begin{array}{c}
I_{1}(\omega)\\
I_{2}(\omega)\end{array}\right)$are the currents measured at the boundaries ($x=-L_{1}$and $x=L_{2}$).
$\left(\begin{array}{c}
I_{T}^{-}(x=0,\omega)\\
I_{T}^{+}(x=0,\omega)\end{array}\right)$ are the currents injected at the STM tip but since we are not in
a DC regime we must take into account propagation effects: that's
why we consider the values taken at the boundaries$\left(\begin{array}{c}
I_{T}^{-}(-L_{1},\omega)\\
I_{T}^{+}(L_{2},\omega)\end{array}\right)$rather than $\left(\begin{array}{c}
I_{T}^{-}(x=0,\omega)\\
I_{T}^{+}(x=0,\omega)\end{array}\right)$ (the components of the two vectors only differ by phases).

If impedance matching is realized clearly:$\left(\begin{array}{c}
I_{1}(\omega)\\
I_{2}(\omega)\end{array}\right)=\left(\begin{array}{c}
I_{T}^{-}(-L_{1},\omega)\\
I_{T}^{+}(L_{2},\omega)\end{array}\right)$.

\subsubsection{AC shot noise: transfer tensor for the current-current correlators
between injected and measured currents.}

Since: \begin{align*}
\left(\begin{array}{c}
I_{1}(\omega)\\
I_{2}(\omega)\end{array}\right) & =\frac{2Z_{0}}{\left(Z_{S}+Z_{0}\right)\left(Z_{D}+Z_{0}\right)\exp-i\left(\phi_{1}+\phi_{2}\right)-\left(Z_{S}-Z_{0}\right)\left(Z_{D}-Z_{0}\right)\exp i\left(\phi_{1}+\phi_{2}\right)}\\
 & \times\left(\begin{array}{cc}
\left(Z_{D}+Z_{0}\right)e^{-i\left(\phi_{1}+\phi_{2}\right)} & \left(Z_{D}-Z_{0}\right)\\
\left(Z_{S}-Z_{0}\right) & \left(Z_{S}+Z_{0}\right)e^{-i\left(\phi_{1}+\phi_{2}\right)}\end{array}\right)\left(\begin{array}{c}
I_{T}^{-}(-L_{1},\omega)\\
I_{T}^{+}(L_{2},\omega)\end{array}\right)\end{align*}
 it follows that: \begin{align*}
\left\langle I_{1}(\omega)I_{1}(-\omega)\right\rangle  & =\frac{\left(2Z_{0}\right)^{2}}{\Delta(\phi_{1}+\phi_{2})\Delta(-\phi_{1}-\phi_{2})}\left\{ \begin{array}{c}
\left(Z_{D}+Z_{0}\right)^{2}S_{-L_{1},-L_{1}}+\left(Z_{D}-Z_{0}\right)^{2}S_{L_{2},L_{2}}\\
+\left(Z_{D}^{2}-Z_{0}^{2}\right)\left[e^{-i\left(\phi_{1}+\phi_{2}\right)}S_{-L_{1},L_{2}}+e^{i\left(\phi_{1}+\phi_{2}\right)}S_{L_{2},-L_{1}}\right]\end{array}\right\} \\
\left\langle I_{2}(\omega)I_{2}(-\omega)\right\rangle  & =\frac{\left(2Z_{0}\right)^{2}}{\Delta(\phi_{1}+\phi_{2})\Delta(-\phi_{1}-\phi_{2})}\left\{ \begin{array}{c}
\left(Z_{S}-Z_{0}\right)^{2}S_{-L_{1},-L_{1}}+\left(Z_{S}+Z_{0}\right)^{2}S_{L_{2},L_{2}}\\
+\left(Z_{S}^{2}-Z_{0}^{2}\right)\left[e^{i\left(\phi_{1}+\phi_{2}\right)}S_{-L_{1},L_{2}}+e^{-i\left(\phi_{1}+\phi_{2}\right)}S_{L_{2},-L_{1}}\right]\end{array}\right\} \\
\left\langle I_{1}(\omega)I_{2}(-\omega)\right\rangle  & =\frac{\left(2Z_{0}\right)^{2}}{\Delta(\phi_{1}+\phi_{2})\Delta(-\phi_{1}-\phi_{2})}\left\{ \begin{array}{c}
\left(Z_{D}+Z_{0}\right)\left(Z_{S}-Z_{0}\right)e^{-i\left(\phi_{1}+\phi_{2}\right)}S_{-L_{1},-L_{1}}\\
+\left(Z_{D}-Z_{0}\right)\left(Z_{S}+Z_{0}\right)e^{i\left(\phi_{1}+\phi_{2}\right)}S_{L_{2},L_{2}}\\
+\left(Z_{D}+Z_{0}\right)\left(Z_{S}+Z_{0}\right)S_{-L_{1},L_{2}}\\
+\left(Z_{S}-Z_{0}\right)\left(Z_{D}-Z_{0}\right)S_{L_{2},-L_{1}}\end{array}\right\} \end{align*}
 where we have defined for convenience: \[
\Delta(\phi_{1}+\phi_{2})=\left(Z_{S}+Z_{0}\right)\left(Z_{D}+Z_{0}\right)\exp-i\left(\phi_{1}+\phi_{2}\right)-\left(Z_{S}-Z_{0}\right)\left(Z_{D}-Z_{0}\right)\exp i\left(\phi_{1}+\phi_{2}\right)\]
 (the denominator) and: \begin{align*}
S_{-L_{1},-L_{1}} & =\left\langle I_{T}^{-}(-L_{1},\omega)I_{T}^{-}(-L_{1},-\omega)\right\rangle \\
S_{L_{2},L_{2}} & =\left\langle I_{T}^{+}(L_{2},\omega)I_{T}^{+}(L_{2},-\omega)\right\rangle \\
S_{-L_{1},L_{2}} & =\left\langle I_{T}^{-}(-L_{1},\omega)I_{T}^{+}(L_{2},-\omega)\right\rangle \\
S_{L_{2},-L_{1}} & =\left\langle I_{T}^{+}(L_{2},\omega)I_{T}^{-}(-L_{1},-\omega)\right\rangle \end{align*}

\subsubsection{Finite frequency dependence of the shot noise with a DC bias: the
case of symetric injection.}

We will for simplicity restrict ourselves to symetric injection of
$+k_{F}$ or $-k_{F}$ electrons by the STM tip subjected to a DC
bias. 

While in the discussion of DC excess noise we were able to compute
the Fano ratios of the injected currents by using the fact that the
injection of particles by the STM is Poissonian no simple calculation
is possible for the AC correlators of $I_{T}^{-}$ and $I_{T}^{+}$.
This stems from the fact that one would need time-resolved information
rather than just the statistics of the total number of particles injected
by the STM which is enough for DC.

An obvious solution would be to make a Keldysh calculation.

Some simple assumptions on $I_{T}^{-}$ and $I_{T}^{+}$ will allow
us to avoid this route while still getting the essential physics:
let us assume that their correlators are identical with those of the
infinite system. According to our discussion of impedance matching
it is likely that their current-current correlators $S_{-L_{1},-L_{1}}$,
$S_{L_{2},L_{2}}$, $S_{-L_{1},L_{2}}$ and $S_{L_{2},-L_{1}}$ are
indeed not too different from those of the infinite LL, which were
computed in \cite{stm} for a four-mode system (a carbon nanotube)
and which, adapted to the spinless LL, read:

\begin{align*}
S_{-L_{1},-L_{1}} & =\left(Q_{+}^{2}+Q_{-}^{2}\right)~\theta(\left|\frac{eV}{\hbar}\right|-\left|\omega\right|)~\left(1-\left|\frac{\hbar\omega}{eV}\right|\right)^{\nu}~I^{\infty}(\omega=0)\\
S_{L_{2},L_{2}} & =S_{-L_{1},-L_{1}}\\
S_{-L_{1},L_{2}} & =2Q_{+}Q_{-}~\theta(\left|\frac{eV}{\hbar}\right|-\left|\omega\right|)~\left(1-\left|\frac{\hbar\omega}{eV}\right|\right)^{\nu}~e^{i\left(\phi_{1}-\phi_{2}\right)}~I^{\infty}(\omega=0)\\
S_{L_{2},-L_{1}} & =2Q_{+}Q_{-}~\theta(\left|\frac{eV}{\hbar}\right|-\left|\omega\right|)~\left(1-\left|\frac{\hbar\omega}{eV}\right|\right)^{\nu}~e^{-i\left(\phi_{1}-\phi_{2}\right)}~I^{\infty}(\omega=0)\end{align*}
 where $\nu=\frac{1}{2}(K+K^{-1})$ and $I^{\infty}(\omega=0)$ is
the DC current injected into one branch:

\[
I^{\infty}(\omega=0)=\frac{2e^{2}\Gamma^{2}}{\pi v_{F}\Gamma(\nu+1)}\left(\frac{a}{u}\right)^{\nu}\left(eV\right)^{\nu}.\]
The direct and cross correlators show (i) the charges of the fractional
states, (ii) and have a characteristic power-law dependence towards
a threshold Josephson frequency. 

We will abbreviate \[
f(V,\omega)=\theta(\left|\frac{eV}{\hbar}\right|-\left|\omega\right|)~\left(1-\left|\frac{\hbar\omega}{eV}\right|\right)^{\nu}~I^{\infty}(\omega=0)\]
 to shorten the lines of algebra.

Using these results one can plug them in the expression found for
the currents in the finite geometry so that: \begin{align*}
\left\langle I_{1}(\omega)I_{1}(-\omega)\right\rangle  & =\frac{\left(2Z_{0}\right)^{2}f(V,\omega)}{\Delta(\phi_{1}+\phi_{2})\Delta(-\phi_{1}-\phi_{2})}\left\{ \begin{array}{c}
\left[\left(Z_{D}+Z_{0}\right)^{2}+\left(Z_{D}-Z_{0}\right)^{2}\right]\left(Q_{+}^{2}+Q_{-}^{2}\right)\\
+4\cos2\phi_{2}\left(Z_{D}^{2}-Z_{0}^{2}\right)Q_{+}Q_{-}\end{array}\right\} \\
\left\langle I_{2}(\omega)I_{2}(-\omega)\right\rangle  & =\frac{\left(2Z_{0}\right)^{2}f(V,\omega)}{\Delta(\phi_{1}+\phi_{2})\Delta(-\phi_{1}-\phi_{2})}\left\{ \begin{array}{c}
\left[\left(Z_{S}+Z_{0}\right)^{2}+\left(Z_{S}-Z_{0}\right)^{2}\right]\left(Q_{+}^{2}+Q_{-}^{2}\right)\\
+4\cos2\phi_{1}\left(Z_{S}^{2}-Z_{0}^{2}\right)Q_{+}Q_{-}\end{array}\right\} \\
\left\langle I_{1}(\omega)I_{2}(-\omega)\right\rangle  & =\frac{\left(2Z_{0}\right)^{2}f(V,\omega)}{\Delta(\phi_{1}+\phi_{2})\Delta(-\phi_{1}-\phi_{2})}\left\{ \begin{array}{c}
\left[\left(Z_{D}+Z_{0}\right)\left(Z_{S}-Z_{0}\right)e^{-i\left(\phi_{1}+\phi_{2}\right)}+\left(Z_{D}-Z_{0}\right)\left(Z_{S}+Z_{0}\right)e^{i\left(\phi_{1}+\phi_{2}\right)}\right]\left(Q_{+}^{2}+Q_{-}^{2}\right)\\
+2Q_{+}Q_{-}\left[\left(Z_{D}+Z_{0}\right)\left(Z_{S}+Z_{0}\right)e^{i\left(\phi_{1}-\phi_{2}\right)}+\left(Z_{S}-Z_{0}\right)\left(Z_{D}-Z_{0}\right)e-^{i\left(\phi_{1}-\phi_{2}\right)}\right]\end{array}\right\} \end{align*}

We can also write:\begin{align*}
\left\langle I_{1}(\omega)I_{1}(-\omega)\right\rangle  & =f(V,\omega)~A~\left[2\left(Z_{D}^{2}+Z_{0}^{2}\right)\left(Q_{+}^{2}+Q_{-}^{2}\right)+4\cos2\phi_{2}\left(Z_{D}^{2}-Z_{0}^{2}\right)Q_{+}Q_{-}\right]\\
 & =f(V,\omega)~2A~\left[Z_{D}^{2}~\left(Q_{+}^{2}+Q_{-}^{2}+2\cos2\phi_{2}~Q_{+}Q_{-}\right)+Z_{0}^{2}~\left(Q_{+}^{2}+Q_{-}^{2}-2\cos2\phi_{2}~Q_{+}Q_{-}\right)\right]\end{align*}
 where: \begin{align*}
A & =\frac{4Z_{0}^{2}}{\left|\Delta(\phi_{1}+\phi_{2})\right|^{2}}\\
 & =\frac{Z_{0}^{2}}{Z_{0}^{2}\left(Z_{S}+Z_{D}\right)^{2}+\sin^{2}(\phi_{1}+\phi_{2})\left(Z_{D}^{2}-Z_{0}^{2}\right)\left(Z_{S}^{2}-Z_{0}^{2}\right)};\end{align*}
 from the previous expression one can see clearly how there can be
a cancellation of the LL parameter $K$ at zero frequency: if $\phi_{2}=0$
$\left(Q_{+}^{2}+Q_{-}^{2}+2\cos2\phi_{2}~Q_{+}Q_{-}\right)$ becomes
$\left(Q_{+}+Q_{-}\right)^{2}=1$ while $Z_{0}^{2}~\left(Q_{+}^{2}+Q_{-}^{2}-2\cos2\phi_{2}~Q_{+}Q_{-}\right)=Z_{0}^{2}~\left(Q_{+}-Q_{-}\right)^{2}=\left(\frac{h}{2e^{2}K}\right)^{2}K^{2}$;
all dependence on $K$ has vanished. The special character of the
non-integer charges of the fractional excitations has disappeared
in the DC limit.

Likewise: \begin{align*}
\left\langle I_{2}(\omega)I_{2}(-\omega)\right\rangle  & =f(V,\omega)~A~\left[2\left(Z_{S}^{2}+Z_{0}^{2}\right)\left(Q_{+}^{2}+Q_{-}^{2}\right)+4\cos2\phi_{1}\left(Z_{S}^{2}-Z_{0}^{2}\right)Q_{+}Q_{-}\right]\\
\left\langle I_{1}(\omega)I_{2}(-\omega)\right\rangle  & =f(V,\omega)~A~\left\{ \begin{array}{c}
\left(Q_{+}^{2}+Q_{-}^{2}\right)\left[\left(Z_{D}+Z_{0}\right)\left(Z_{S}-Z_{0}\right)e^{-i(\phi_{1}+\phi_{2})}+\left(Z_{D}-Z_{0}\right)\left(Z_{S}+Z_{0}\right)e^{i(\phi_{1}+\phi_{2})}\right]\\
+2Q_{+}Q_{-}\left[\left(Z_{D}+Z_{0}\right)\left(Z_{S}+Z_{0}\right)e^{i(\phi_{1}-\phi_{2})}+\left(Z_{D}-Z_{0}\right)\left(Z_{S}-Z_{0}\right)e^{-i(\phi_{1}-\phi_{2})}\right]\end{array}\right\} \end{align*}
where $\phi_{2}=\frac{\omega}{u}L_{2}$ and $\phi_{1}=\frac{\omega}{u}L_{1}$
($L_{1}$ and $L_{2}$ are the lengths from the impurity to each boundary).
This is the main result of this paper.

It is easy although tedious to check that the expressions derived
in \cite{stm} for the finite LL correspond to the limiting case:
$Z_{S}=Z_{D}=h/2e^{2}$. Our expression has the same range of validity
as theirs: it yields the dominant length-dependent oscillating contribution
to the noise. 

We observe in passing that our expressions are valid only for excess
noise (not equilibrium noise): this stems from the fact that we have
used the expression of the excess noise of the injected currents.

The Fano ratios (correlators divided by currents) follow straightforwardly:
to make progress we still assume that the injected currents will be
identical with the currents of the infinite system, which should be
correct to leading order in $1/L$ the length of the system: $\left\langle I_{T}^{-}\right\rangle =\left\langle I_{1}^{\infty}\right\rangle $.
Since $I_{1}=\frac{Z_{D}+Z_{0}}{Z_{S}+Z_{D}}I_{T}^{-}+\frac{Z_{D}-Z_{0}}{Z_{S}+Z_{D}}I_{T}^{+}$
one eventually finds (since injection is symetric):\begin{align*}
\left\langle I_{1}\right\rangle  & =\frac{2Z_{D}}{Z_{S}+Z_{D}}\left\langle I_{T}^{-}\right\rangle \\
 & =\frac{2Z_{D}}{Z_{S}+Z_{D}}\left\langle I^{\infty}\right\rangle \end{align*}
 So that: \begin{align*}
\frac{\left\langle I_{1}(\omega)I_{1}(-\omega)\right\rangle }{\left\langle I_{1}(\omega=0)\right\rangle } & =\theta(\left|eV\right|-\left|\hbar\omega\right|)~\left(1-\left|\frac{\hbar\omega}{eV}\right|\right)^{\nu}~\frac{Z_{S}+Z_{D}}{2Z_{D}}~\\
 & \times A~\left[2\left(Z_{D}^{2}+Z_{0}^{2}\right)\left(Q_{+}^{2}+Q_{-}^{2}\right)+4\cos2\phi_{2}\left(Z_{D}^{2}-Z_{0}^{2}\right)Q_{+}Q_{-}\right]\\
 & =\frac{1}{2}\theta(\left|eV\right|-\left|\hbar\omega\right|)~\left(1-\left|\frac{\hbar\omega}{eV}\right|\right)^{\nu}~\frac{Z_{S}+Z_{D}}{2Z_{D}}~\\
 & \times\frac{Z_{0}^{2}~\left[2\left(Z_{D}^{2}+Z_{0}^{2}\right)\left(Q_{+}^{2}+Q_{-}^{2}\right)+4\cos2\phi_{2}\left(Z_{D}^{2}-Z_{0}^{2}\right)Q_{+}Q_{-}\right]}{Z_{0}^{2}\left(Z_{S}+Z_{D}\right)^{2}+\sin^{2}(\phi_{1}+\phi_{2})\left(Z_{D}^{2}-Z_{0}^{2}\right)\left(Z_{S}^{2}-Z_{0}^{2}\right)}\end{align*}
and: \begin{align*}
\frac{\left\langle I_{1}(\omega)I_{2}(-\omega)\right\rangle }{\left\langle I_{1}(\omega=0)\right\rangle } & =\theta(\left|eV\right|-\left|\hbar\omega\right|)~\left(1-\left|\frac{\hbar\omega}{eV}\right|\right)^{\nu}~\frac{Z_{S}+Z_{D}}{2Z_{D}}~\\
 & \times\frac{Z_{0}^{2}~}{Z_{0}^{2}\left(Z_{S}+Z_{D}\right)^{2}+\sin^{2}(\phi_{1}+\phi_{2})\left(Z_{D}^{2}-Z_{0}^{2}\right)\left(Z_{S}^{2}-Z_{0}^{2}\right)}\\
 & \times\left\{ \begin{array}{c}
\left(Q_{+}^{2}+Q_{-}^{2}\right)\left[\left(Z_{D}+Z_{0}\right)\left(Z_{S}-Z_{0}\right)e^{-i(\phi_{1}+\phi_{2})}+\left(Z_{D}-Z_{0}\right)\left(Z_{S}+Z_{0}\right)e^{i(\phi_{1}+\phi_{2})}\right]\\
+2Q_{+}Q_{-}\left[\left(Z_{D}+Z_{0}\right)\left(Z_{S}+Z_{0}\right)e^{i(\phi_{1}-\phi_{2})}+\left(Z_{D}-Z_{0}\right)\left(Z_{S}-Z_{0}\right)e^{-i(\phi_{1}-\phi_{2})}\right]\end{array}\right\} \end{align*}
This yields ratios which are independent of the exact variations of
the DC currents, which can be advantageous.

\subsubsection{Discussion: experimental implications.}

Lebedev et al. \cite{stm} find the following expressions for the
excess noise, corresponding to $Z_{S}=Z_{D}=h/2e^{2}$:\begin{equation}
\left\langle I(x,\omega)I(x',-\omega)\right\rangle =\frac{1}{2}f(V,\omega)~\left[\frac{1}{1-(1-K^{-2})\sin^{2}\phi}+\frac{sgn(x)sgn(x')}{1-(1-K^{2})\sin^{2}\phi}\right]\label{eq:stm-noise}\end{equation}
for $x=\pm x'=\pm L$ and with currents oriented as outgoing from
the tunneling point ($x=0$).

A issue with these expressions is that the LL parameter $K$ plays
a dual role: it intervenes in the charges of the elementary fractional
excitations and also it gives the characteristic impedance of the
system which regulates reflections. The Fano factor mixes both contributions
and therefore the previous expression conceal the charges: if used
experimentally such equations can only provide a measurement of the
LL parameter $K$; they do not show clearly how shot noise measures
the charges. 

This is contrast with our approach where the leading (oscillating)
contribution to the Fano factors coming from the reflections have
explictily been built. This explains why we are able to sort out the
contribution coming from the fractional charges: \begin{align}
\left\langle I_{1}(\omega)I_{1}(-\omega)\right\rangle  & =f(V,\omega)~A~\left[2\left(Z_{D}^{2}+Z_{0}^{2}\right)\left(Q_{+}^{2}+Q_{-}^{2}\right)+4\cos2\phi_{2}\left(Z_{D}^{2}-Z_{0}^{2}\right)Q_{+}Q_{-}\right]\\
\left\langle I_{2}(\omega)I_{2}(-\omega)\right\rangle  & =f(V,\omega)~A~\left[2\left(Z_{S}^{2}+Z_{0}^{2}\right)\left(Q_{+}^{2}+Q_{-}^{2}\right)+4\cos2\phi_{1}\left(Z_{S}^{2}-Z_{0}^{2}\right)Q_{+}Q_{-}\right]\\
\left\langle I_{1}(\omega)I_{2}(-\omega)\right\rangle  & =f(V,\omega)~A~\left\{ \begin{array}{c}
\left(Q_{+}^{2}+Q_{-}^{2}\right)\left[\left(Z_{D}+Z_{0}\right)\left(Z_{S}-Z_{0}\right)e^{-i(\phi_{1}+\phi_{2})}+\left(Z_{D}-Z_{0}\right)\left(Z_{S}+Z_{0}\right)e^{i(\phi_{1}+\phi_{2})}\right]\\
+2Q_{+}Q_{-}\left[\left(Z_{D}+Z_{0}\right)\left(Z_{S}+Z_{0}\right)e^{i(\phi_{1}-\phi_{2})}+\left(Z_{D}-Z_{0}\right)\left(Z_{S}-Z_{0}\right)e^{-i(\phi_{1}-\phi_{2})}\right]\end{array}\right\} \label{eq:stm-new}\end{align}

We also stress that our expressions do not depend on the explicit
values of the fractional charges, which means that these formulas
can be used experimentally without making any a priori assumption
on $Q_{\pm}$.

Once one has independent values of the characteristic impedance $Z_{0}$
and of the interface resistances $Z_{S}$ and $Z_{D}$ (through for
instance AC conductance measurements) the shot noise allows unambiguous
extraction -without any fitting parameter- of the charges $Q_{+}$
and $Q_{-}$. As a bonus we have then a distinct LL theory prediction
which can then be further checked, namely: $Q_{\pm}=\frac{1\pm K}{2}=\left(1\pm\frac{h}{2e^{2}}Z_{0}^{-1}\right)/2$
which is evidence of the fractionalization of the electron.

There are several strategies for using these excess noise correlators
experimentally to extract fractional charges: 

(i) Lebedev, Crépieux and Martin \cite{stm} propose to measure ratios
of cross and direct correlations at a \textit{resonance frequency}:
this demands being able to make probes at rather large frequencies
(at least $1-100\: GHz$ ). 

(ii) Since one has the exact dependence of the noise on the fractional
charges it is a better strategy to use these relations directly by
measuring the deviations to the DC limit at lower frequencies: for
instance for a $1$\% deviation this lowers the frequency range by
a factor of ten (for the direct correlators) or even a hundred (for
the crossed correlators) to $10\: MHz-10\: GHz$. This is because
the cross-correlations have a linear term in $\omega$ in a low-frequency
expansion while the direct correlations go as $\omega^{2}$:\[
\frac{\left\langle I_{1}(\omega)I_{2}(-\omega)\right\rangle }{\left\langle I_{1}(\omega)I_{1}(-\omega)\right\rangle }=\frac{Z_{S}Z_{D}-R_{0}^{2}/4}{Z_{D}^{2}+R_{0}^{2}/4}+i\frac{\omega}{u}\frac{2Z_{0}}{Z_{D}^{2}+R_{0}^{2}/4}\left\{ \left(L_{1}+L_{2}\right)\left(Z_{D}-Z_{S}\right)\left(Q_{+}^{2}+Q_{-}^{2}\right)+\left(L_{1}-L_{2}\right)\left(Z_{D}+Z_{S}\right)\left(2Q_{+}Q_{-}\right)\right\} \]
Observe that in order to extract both the factors $Q_{+}^{2}+Q_{-}^{2}$
and $Q_{+}Q_{-}$one needs both $Z_{D}\neq Z_{S}$ and $L_{1}\neq L_{2}$.
\emph{The symetric geometry is therefore the least favorable to observe
the deviations to DC excess noise}.

(iii) The frequency range is improved but remains still high. For
that reason, it seems much better to make an impedance matching which
will already yield the fractional charges at the DC range. This is
the strategy we advocate.

\section{Backscattering by an impurity.}

In this section we discuss the topic of a single impurity in a finite
Luttinger liquid connected to reservoirs.%
\begin{figure}[h]
\includegraphics{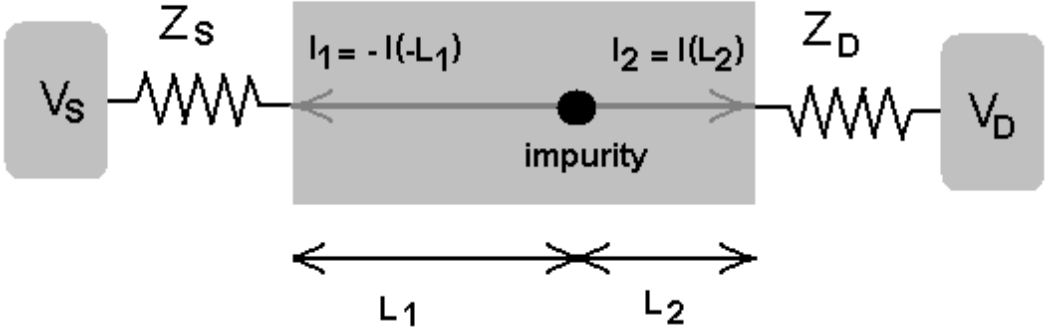}

\caption{\label{cap:impurity-setting}LL with an impurity connected to impedances
(as boundary conditions).}
\end{figure}

\subsection{Reduction to the STM problem.}

While the Keldysh approach can be used we propose another physically
more transparent method which relies on a fine examination of current
operators and impedance mismatch, much in the spirit of our treatment
of the STM problem. Additionally our method gives access to excess
noise but is not plagued with the ambiguities created by the involvement
of the LL parameter $K$ in both the fractional charge and the reflection
coefficients: the expresssions for the excess noise found through
the Keldysh approach would involve the parameter $K$ without differentitating

The idea again is to reduce the current operators to another set of
current operators, whose correlators are easier to compute. The new
set of operators corresponds physically to currents in an impedance-matched
environment. 

\begin{figure}
\includegraphics{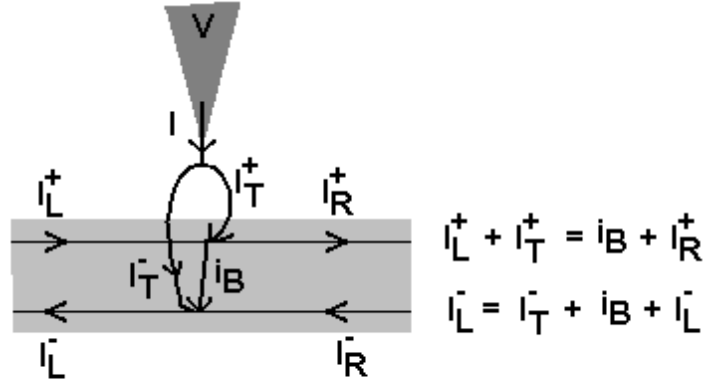}

\caption{\label{cap:def-of-currents}Definition of the chiral currents in
the most general setting including both backscattering by an impurity
and tunneling current from a STM tip. Note however that in this section
the discussion is specialized to sole impurity backscattering so that
$I_{T}^{+}=0$ and $I_{T}^{-}=0$ (the STM tip is removed). }
\end{figure}

So let us consider $i_{B}$ \emph{the current backscattered from one
chiral (plasma) branch to the other}: it is NOT the electronic backscattering
current which is simply (to the right of the impurity) $I_{2}-I_{0}$the
difference between the current in the presence of the impurity and
the current in the absence of an impurity. This follows from their
expression; in a general setting where one has both backscattering
and current injection one would have (see Figure \ref{cap:def-of-currents}):\begin{eqnarray*}
i_{B}-I_{T}^{+} & = & i_{L}^{+}-i_{R}^{+}\\
i_{B}+I_{T}^{-} & = & i_{R}^{-}-i_{L}^{-}\end{eqnarray*}
which reduces here to:\[
i_{B}=i_{R}^{-}-i_{L}^{-}=i_{L}^{+}-i_{R}^{+}\]
by current conservation. 

In contrast:\[
I_{0}-I_{2}=\frac{e^{2}}{h}V_{SD}-(i_{R}^{+}-i_{R}^{-}).\]
 ($V_{SD}$ is the voltage between source and drain). It is a simple
matter however to show that in the infinite system $\left\langle I_{2}-I_{0}\right\rangle =\left\langle i_{B}\right\rangle $.
We might say that $i_{B}$ gives information about the backscattering
at the impurity while $I_{2}-I_{0}$ contains the full backscattering,
including the reflections at boundaries.

We turn to the boundary conditions which now include the voltage of
source and drain: \begin{align*}
Z_{S}I_{1} & =\left[\frac{\left(\mu_{L}^{+}+\mu_{L}^{-}\right)}{2e}-V_{S}\right]=Z_{0}\left(i_{L}^{+}+i_{L}^{-}\right)-V_{S}\\
Z_{D}I_{2} & =\left[\frac{\left(\mu_{R}^{+}+\mu_{R}^{-}\right)}{2e}-V_{D}\right]=Z_{0}\left(i_{R}^{+}+i_{R}^{-}\right)-V_{D}\end{align*}
 Thus : \[
Z_{S}I_{1}-Z_{D}I_{2}=Z_{0}2i_{B}+V_{D}-V_{S}\]
while: \[
I_{1}+I_{2}=0\]
 (by current conservation along the wire). 

We define observables $I_{1}^{0}$ and $I_{2}^{0}$ in the absence
of the impurity $i_{B}=0$ ; they obey \begin{align*}
Z_{S}I_{1}^{0}-Z_{D}I_{2}^{0} & =V_{D}-V_{S}\\
I_{1}^{0}+I_{2}^{0} & =0\end{align*}
 Then for the shifted variables $\delta I_{1}=I_{1}-I_{1}^{0}$ and
$\delta I_{2}=I_{2}-I_{2}^{0}$ we get: \begin{align*}
\delta I_{1}+\delta I_{2} & =0\\
Z_{S}\delta I_{1}-Z_{D}\delta I_{2} & =Z_{0}2i_{B}\end{align*}
 The first line simply expresses the fact that the same current is
backscattered to the right and to the left of impurity. This is the
same set of equations we got with the injected current operators $I_{T}^{-}$
and $I_{T}^{+}$ (see Eq. (\ref{cur-ren}) ) if we identify $I_{T}^{+}=-I_{T}^{-}=i_{B}$
with $\delta I_{1}+\delta I_{2}=0$. We therefore arrive at the main
point of the section, namely that the same matrix equation holds:
\[
\left(\begin{array}{c}
\delta I_{1}\\
\delta I_{2}\end{array}\right)=\frac{1}{Z_{S}+Z_{D}}\left(\begin{array}{cc}
Z_{D}+Z_{0} & Z_{D}-Z_{0}\\
Z_{S}-Z_{0} & Z_{S}+Z_{0}\end{array}\right)\left(\begin{array}{c}
i_{B}\\
-i_{B}\end{array}\right).\]

It follows that this equation admits the same interpretation as for
the STM tip: the current measured in the leads and the current backscattered
by the impurity are different objects; we can view the above equation
as an \emph{operator renormalization of the backscattering current
operator}, which results from multiple reflections at the boundaries
of the system.

\subsection{Finite-frequency.}

The previous equations for the tunneling currents at finite frequency
\begin{align*}
\left(\begin{array}{c}
I_{1}\\
I_{2}\end{array}\right) & =\frac{2Z_{0}}{\left(Z_{S}+Z_{0}\right)\left(Z_{D}+Z_{0}\right)\exp i\left(\phi_{1}+\phi_{2}\right)-\left(Z_{S}-Z_{0}\right)\left(Z_{D}-Z_{0}\right)\exp-i\left(\phi_{1}+\phi_{2}\right)}\\
 & \times\left(\begin{array}{cc}
\left(Z_{D}+Z_{0}\right)e^{-i\phi_{2}} & \left(Z_{D}-Z_{0}\right)e^{i\phi_{2}}\\
\left(Z_{S}-Z_{0}\right)e^{i\phi_{1}} & \left(Z_{S}+Z_{0}\right)e^{-i\phi_{1}}\end{array}\right)\left(\begin{array}{c}
I_{T}^{-}\\
I_{T}^{+}\end{array}\right)\end{align*}
 are modified like this: \begin{align*}
\left(\begin{array}{c}
\delta I_{1}\\
\delta I_{2}\end{array}\right) & =\frac{2Z_{0}\quad i_{B}}{\left(Z_{S}+Z_{0}\right)\left(Z_{D}+Z_{0}\right)\exp i\left(\phi_{1}+\phi_{2}\right)-\left(Z_{S}-Z_{0}\right)\left(Z_{D}-Z_{0}\right)\exp-i\left(\phi_{1}+\phi_{2}\right)}\\
 & \times\left(\begin{array}{c}
\left(Z_{D}+Z_{0}\right)e^{-i\phi_{2}}-\left(Z_{D}-Z_{0}\right)e^{i\phi_{2}}\\
\left(Z_{S}-Z_{0}\right)e^{i\phi_{1}}-\left(Z_{S}+Z_{0}\right)e^{-i\phi_{1}}\end{array}\right)\end{align*}

Observe that at finite frequency that $\delta I_{1}+\delta I_{2}\neq0$
(and that $I_{1}+I_{2}\neq0$ : there is a charging of the system).

\subsection{Reduction of the current correlators to simpler ones.}

From the relation between the measured currents and the backscattering
current at finite frequency one finds that the currents in a matched
geometry obey: \begin{align*}
\left\langle \delta I_{1}^{matched}\right\rangle  & =e^{i\phi_{1}}\left\langle i_{B}\right\rangle \\
\left\langle \delta I_{2}^{matched}\right\rangle  & =e^{i\phi_{2}}\left\langle i_{B}\right\rangle \end{align*}
 It is therefore convenient to define the following operators since
for the matched geometry they will have excpectation values identical
to those of the infinite system:\begin{align*}
\delta I'_{1} & =e^{i\phi_{1}}i_{B}\\
\delta I'_{2} & =e^{i\phi_{2}}i_{B}\end{align*}
So that at finite frequency: \begin{align*}
\delta I_{1} & =\frac{2Z_{0}}{\Delta\left(\phi_{1}+\phi_{2}\right)}~e^{-i\phi_{1}}~\left[\left(Z_{D}+Z_{0}\right)e^{-i\phi_{2}}-\left(Z_{D}-Z_{0}\right)e^{i\phi_{2}}\right]~\delta I'_{1}\\
\delta I_{2} & =\frac{2Z_{0}}{\Delta\left(\phi_{1}+\phi_{2}\right)}~e^{-i\phi_{2}}~\left[\left(Z_{S}-Z_{0}\right)e^{i\phi_{1}}-\left(Z_{S}+Z_{0}\right)e^{-i\phi_{1}}\right]~\delta I'_{2}\end{align*}
 (and at zero frequency: $\delta I_{1}=\frac{2Z_{0}}{Z_{S}+Z_{D}}~\delta I'_{1}$).

Finally the relation between the correlators of matched and unmatched
system are: \begin{eqnarray*}
\left\langle \delta I_{1}(\omega)\delta I_{1}(-\omega)\right\rangle  & = & 4~A~\left(Z_{D}^{2}\sin^{2}\phi_{2}+Z_{0}^{2}\cos^{2}\phi_{2}\right)~\left\langle \delta I'_{1}(\omega)\delta I'_{1}(-\omega)\right\rangle \\
\frac{\left\langle \delta I_{1}(\omega)\delta I_{1}(-\omega)\right\rangle }{\left\langle \delta I'_{1}(\omega)\delta I'_{1}(-\omega)\right\rangle } & = & \frac{4Z_{0}^{2}~\left(Z_{D}^{2}\sin^{2}\phi_{2}+Z_{0}^{2}\cos^{2}\phi_{2}\right)}{Z_{0}^{2}\left(Z_{S}+Z_{D}\right)^{2}+\sin^{2}(\phi_{1}+\phi_{2})\left(Z_{D}^{2}-Z_{0}^{2}\right)\left(Z_{S}^{2}-Z_{0}^{2}\right)}\\
\frac{\left\langle \delta I_{2}(\omega)\delta I_{2}(-\omega)\right\rangle }{\left\langle \delta I'_{2}(\omega)\delta I'_{2}(-\omega)\right\rangle } & = & \frac{4Z_{0}^{2}~\left(Z_{S}^{2}\sin^{2}\phi_{1}+Z_{0}^{2}\cos^{2}\phi_{1}\right)}{Z_{0}^{2}\left(Z_{S}+Z_{D}\right)^{2}+\sin^{2}(\phi_{1}+\phi_{2})\left(Z_{D}^{2}-Z_{0}^{2}\right)\left(Z_{S}^{2}-Z_{0}^{2}\right)}\end{eqnarray*}
 This is the main result of this section: the relation is valid at
any temperature, voltage and frequency so long as the LL theory is
valid. Note that the renormalization is temperature independent: it
comes solely from the physics of impedance mismatch. We stress that
this is a non-perturbative relation which must be obeyed by any consistent
theory.

\subsection{Excess noise.}

We now make the following simplifying assumption, as in the STM problem,
namely we approximate the correlators $\left\langle \delta I'_{1}(\omega)\delta I'_{1}(-\omega)\right\rangle $
and $\left\langle \delta I'_{2}(\omega)\delta I'_{2}(-\omega)\right\rangle $
by their values taken in the infinite system: the rationale for doing
this is that again these operators correspond to the impedance matched
current operators.

We plug in the expression of the shot noise in the infinite system:
\[
\left\langle \delta I_{1}^{\infty}(\omega)\delta I_{1}^{\infty}(-\omega)\right\rangle =-Q_{\ast}~\left[\left(1-\left|\frac{\hbar\omega}{eV}\right|\right)^{\alpha}+\left(1+\left|\frac{\hbar\omega}{eV}\right|\right)^{\alpha}\right]~\left\langle \delta I_{1}^{\infty}(\omega=0)\right\rangle \]
 where $Q_{\ast}=K$ is the charge of Laughlin excitations and $\alpha=2K-1$
(see \cite{chamon}).

Finally: \begin{align*}
\frac{\left\langle \delta I_{1}(\omega)\delta I_{1}(-\omega)\right\rangle }{\left\langle \delta I_{1}(\omega=0)\right\rangle } & =-Q_{\ast}~\left[\left(1-\left|\frac{\hbar\omega}{eV}\right|\right)^{\alpha}+\left(1+\left|\frac{\hbar\omega}{eV}\right|\right)^{\alpha}\right]~4~A~\left(Z_{D}^{2}\sin^{2}\phi_{2}+Z_{0}^{2}\cos^{2}\phi_{2}\right)~\frac{Z_{S}+Z_{D}}{2Z_{0}}\\
 & =-Q_{\ast}~\left[\left(1-\left|\frac{\hbar\omega}{eV}\right|\right)^{\alpha}+\left(1+\left|\frac{\hbar\omega}{eV}\right|\right)^{\alpha}\right]\frac{2Z_{0}~\left(Z_{D}^{2}\sin^{2}\phi_{2}+Z_{0}^{2}\cos^{2}\phi_{2}\right)~\left(Z_{S}+Z_{D}\right)}{Z_{0}^{2}\left(Z_{S}+Z_{D}\right)^{2}+\sin^{2}(\phi_{1}+\phi_{2})\left(Z_{D}^{2}-Z_{0}^{2}\right)\left(Z_{S}^{2}-Z_{0}^{2}\right)}\end{align*}
where $\phi_{2}=\frac{\omega}{u}L_{2}$ and $\phi_{1}=\frac{\omega}{u}L_{1}$
($L_{1}$ and $L_{2}$ are the lengths from the impurity to each boundary).
This is one of the main results of the paper. 

Observe that as a subcase of this formula one gets for $Z_{S}=Z_{D}=h/2e^{2}$
(and neglecting the small prefactors): 

\[
\left\langle \delta I_{1}(\omega)\delta I_{1}(-\omega)\right\rangle =-\delta I_{1}~\left(1-\gamma\right)^{2}\frac{1+\gamma^{2}+2\gamma\cos2\phi_{2}}{1+\gamma^{4}-2\gamma^{2}\cos2(\phi_{1}+\phi_{2})}\]
 where $\gamma=\frac{1-K}{1+K}$. This is the expression found using
the Keldysh technique by Dolcini et al \cite{impurity} (the correspondence
to the notations of that paper is that our$\delta I_{1}=-I_{B}$ in
their notations).

\subsection{Discussion.}

Our general formula can be interpreted as follows; it has three main
components:

(i) the anomalous charge $Q_{\ast}$ of the Laughlin excitations;

(ii) a frequency and voltage dependent part which already exists for
the infinite LL;

(iii) the third factor is a renormalization.

As a summary this formula is superiour to those existing in the litterature
for the following reasons:

(i) it is valid for arbitrary values of the load impedances $Z_{S}$
and $Z_{D}$ at source and drain while other expressions in the litterature
\cite{stm,impurity} (to the author's konwledge) are only correct
for $Z_{S}=Z_{D}=h/2e^{2}$.

(ii) Other expressions mix two very distinct aspects of the LL parameter
$K$: as a characteristic impedance and as a charge of the Laughlin
fractional quasiparticles. Indeed they are expressed solely in terms
of the parameter $K$: $\frac{\left\langle \delta I_{1}(\omega)\delta I_{1}(-\omega)\right\rangle }{\left\langle \delta I_{1}(\omega=0)\right\rangle }=f(K)$.
As a result it may not be conceptually clear whether one measures
a fractional charge or just the plain LL parameter $K$. In contrast
our expression \emph{proves} that the noise assumes the simple form
$\frac{\left\langle \delta I_{1}(\omega)\delta I_{1}(-\omega)\right\rangle }{\left\langle \delta I_{1}(\omega=0)\right\rangle }=Q_{\ast}g(K)$
(which is a priori unexpected). The formula is actually valid independently
of the value taken by the fractional charge: it only relies on the
assumption of Poisson scattering of fractional excitations carrying
a charge $Q_{\ast}$. For Poisson scattering this shows the charge
MUST appear as a prefactor and never enters in the renormalizing factor.

For instance if we blur the distinction between $K$, $Q_{\ast}=Ke$
and $Z_{0}=h/2e^{2}K$ we might be tempted to argue that since $Z_{0}^{2}$
appears as a factor of $\cos^{2}\phi_{2}$ in the numerator of the
expression, measuring the prefactor of this term relative to that
of $\sin^{2}\phi_{2}$ provides a measurement of the inverse square
of the anomalous charge $Q_{\ast}=Ke$ : but this is conceptually
completely wrong!

(iii) It has been suggested \cite{impurity} that the integral of
the Fano factor over a period $\omega_{0}=2\pi u/L$ yields the charge
$Q_{\ast}$. However firstly, this is a coincidence for the very special
case where $Z_{S}=Z_{D}=h/2e^{2}$ and secondly this comes about by
neglecting the term $\left[\left(1-\left|\frac{\hbar\omega}{eV}\right|\right)^{\alpha}+\left(1+\left|\frac{\hbar\omega}{eV}\right|\right)^{\alpha}\right]$:
at any rate even if we discard the power-laws in general, the value
of that integral is \[
Q_{\ast}~\frac{Z_{0}^{2}+Z_{D}^{2}}{Z_{0}^{2}+Z_{S}Z_{D}}.\]
 So it is only by accident that one finds the fractional charge $Q_{\ast}$
when $Z_{S}=Z_{D}=h/2e^{2}$. It is quite unlikely that one might
have both identical electrodes AND an impurity sitting exactly at
the middle of the wire ($L_{1}=L_{2}$), which is one of the conditions
under which the integral has been computed. So the idea that averaging
the Fano factor over a period yields the fractional charge is in general
incorrect and becomes correct only under some drastic conditions. 

If the left and right contact resistances differ, the additional factor
$\frac{Z_{0}^{2}+Z_{D}^{2}}{Z_{0}^{2}+Z_{S}Z_{D}}$ enters and can
not be discarded. If one uses our more general result for the integral
over a period to extract the charge there is still the drawback that
one must aim at the $1-100\: GHz$ range, which is extremely high.

(iv) In contrast and as in the STM geometry since we have expressions
depending explictily on the fractional charge we can use them at lower
frequencies by measuring deviations to the DC limit, i.e. already
in the range of $100\: MHz-10\: GHz$ for a 1\% variation. The fractional
charge will show as a ratio independent of frequency:\[
Q_{\ast}=-\frac{\left\langle \delta I_{1}(\omega)\delta I_{1}(-\omega)\right\rangle }{\left\langle \delta I_{1}(\omega=0)\right\rangle }\:\frac{Z_{0}^{2}\left(Z_{S}+Z_{D}\right)^{2}+\sin^{2}(\phi_{1}+\phi_{2})\left(Z_{D}^{2}-Z_{0}^{2}\right)\left(Z_{S}^{2}-Z_{0}^{2}\right)}{\left[2Z_{0}~\left(Z_{D}^{2}\sin^{2}\phi_{2}+Z_{0}^{2}\cos^{2}\phi_{2}\right)~\left(Z_{S}+Z_{D}\right)\right]\left[\left(1-\left|\frac{\hbar\omega}{eV}\right|\right)^{\alpha}+\left(1+\left|\frac{\hbar\omega}{eV}\right|\right)^{\alpha}\right]}.\]

(v) Still this remains high. So the method we advocate is again to
match impedances at the boundaries of the setup.

\section{Conclusion and experimental prospects.}

We summarize our results.

1. We have computed the DC and AC excess noise of a Luttinger liquid
connected to two electrodes using a boundary-conditions formalism
describing \textit{interface resistances} connected to the system.
All the expressions for the DC and AC excess shot noise derived in
this paper improve on the existing litterature by separating clearly
the contributions from the charge of the elementary excitations and
the contributions arising from reflections. Both are mixed in other
theories because the charges and the reflections both depend on the
LL parameter $K$. In contrast our derivations rely on a close analysis
of the reflections which enabled us to pinpoint exactly how each factor
enters in the shot noise formulas. Additionally our analysis is quite
simple and does not need the sophisticated machinery of the Keldysh
technique. 

2. For the first experimental setup considered (injection of electrons
by a STM tip within the bulk of a LL) we found that although (because
of reflections at the boundaries) DC shot noise is in general unable
to yield information about the fractional charges carried by the elementary
excitations of the Luttinger liquid, still by using \emph{impedance
matching} one can recover the fractional charges. For such an impedance
matched LL:\begin{align}
\frac{\left\langle \Delta I_{1,matched}^{2}\right\rangle }{\left\langle \Delta I_{1,matched}\right\rangle } & =\frac{Q_{+}^{2}T+Q_{-}^{2}R}{Q_{+}T+Q_{-}R},\\
\frac{\left\langle \Delta I_{2,matched}^{2}\right\rangle }{\left\langle \Delta I_{2,matched}\right\rangle } & =\frac{Q_{-}^{2}T+Q_{+}^{2}R}{Q_{-}T+Q_{+}R},\\
\frac{\left\langle \Delta I_{1,matched}\Delta I_{2,matched}\right\rangle }{\left\langle \Delta I_{1,matched}\right\rangle } & =\frac{Q_{+}Q_{-}}{Q_{+}T+Q_{-}R}.\end{align}
which slightly generalizes the expression found in \cite{stm} by
allowing asymetrical injection (to either the right or left Fermi
points $k_{F}$ and $-k_{F}$). The predictions of \cite{stm} for
the infinite system should therefore be observable although the system
is finite. We note also that our formulas have been established in
this paper independently of the exact values of the charges $Q_{\pm}$.
The LL theory predicts however that: $Q_{+}=\frac{1+K}{2},~Q_{-}=\frac{1-K}{2}$. 

If impedance matching is difficult to realize one can still recover
the fractional charges through measuring \textit{AC shot noise:}\begin{align*}
\left\langle I_{1}(\omega)I_{1}(-\omega)\right\rangle  & =f(V,\omega)~A~\left[2\left(Z_{D}^{2}+Z_{0}^{2}\right)\left(Q_{+}^{2}+Q_{-}^{2}\right)+4\cos2\phi_{2}\left(Z_{D}^{2}-Z_{0}^{2}\right)Q_{+}Q_{-}\right]\\
\left\langle I_{2}(\omega)I_{2}(-\omega)\right\rangle  & =f(V,\omega)~A~\left[2\left(Z_{S}^{2}+Z_{0}^{2}\right)\left(Q_{+}^{2}+Q_{-}^{2}\right)+4\cos2\phi_{1}\left(Z_{S}^{2}-Z_{0}^{2}\right)Q_{+}Q_{-}\right]\\
\left\langle I_{1}(\omega)I_{2}(-\omega)\right\rangle  & =f(V,\omega)~A~\left\{ \begin{array}{c}
\left(Q_{+}^{2}+Q_{-}^{2}\right)\left[\left(Z_{D}+Z_{0}\right)\left(Z_{S}-Z_{0}\right)e^{-i(\phi_{1}+\phi_{2})}+\left(Z_{D}-Z_{0}\right)\left(Z_{S}+Z_{0}\right)e^{i(\phi_{1}+\phi_{2})}\right]\\
+2Q_{+}Q_{-}\left[\left(Z_{D}+Z_{0}\right)\left(Z_{S}+Z_{0}\right)e^{i(\phi_{1}-\phi_{2})}+\left(Z_{D}-Z_{0}\right)\left(Z_{S}-Z_{0}\right)e^{-i(\phi_{1}-\phi_{2})}\right]\end{array}\right\} \end{align*}
where $\phi_{2}=\frac{\omega}{u}L_{2}$ and $\phi_{1}=\frac{\omega}{u}L_{1}$
($L_{1}$ and $L_{2}$ are the lengths from the impurity to each boundary)
and $f(V,\omega)~A=\frac{Z_{0}^{2}}{Z_{0}^{2}\left(Z_{S}+Z_{D}\right)^{2}+\sin^{2}(\phi_{1}+\phi_{2})\left(Z_{D}^{2}-Z_{0}^{2}\right)\left(Z_{S}^{2}-Z_{0}^{2}\right)}~\theta(\left|\frac{eV}{\hbar}\right|-\left|\omega\right|)~\left(1-\left|\frac{\hbar\omega}{eV}\right|\right)^{\nu}~\frac{2e^{2}\Gamma^{2}}{\pi v_{F}\Gamma(\nu+1)}\left(\frac{a}{u}\right)^{\nu}\left(eV\right)^{\nu}$.
This should be in the range of $100\: MHz-10\: GHz$, which improves
by a factor of ten other experimental proposals which rely on the
periodic nature of the noise (as frequency is varied) \cite{impurity}.

3. Similar conclusions apply to the setup consisting of an impurity
sitting in the bulk of a Luttinger liquid. We found that the best
method is still to match impedances although fractional charges should
show with AC probes as:

\[
Q_{\ast}=-\frac{\left\langle \delta I_{1}(\omega)\delta I_{1}(-\omega)\right\rangle }{\left\langle \delta I_{1}(\omega=0)\right\rangle }\:\frac{Z_{0}^{2}\left(Z_{S}+Z_{D}\right)^{2}+\sin^{2}(\phi_{1}+\phi_{2})\left(Z_{D}^{2}-Z_{0}^{2}\right)\left(Z_{S}^{2}-Z_{0}^{2}\right)}{\left[2Z_{0}~\left(Z_{D}^{2}\sin^{2}\phi_{2}+Z_{0}^{2}\cos^{2}\phi_{2}\right)~\left(Z_{S}+Z_{D}\right)\right]\left[\left(1-\left|\frac{\hbar\omega}{eV}\right|\right)^{\alpha}+\left(1+\left|\frac{\hbar\omega}{eV}\right|\right)^{\alpha}\right]}.\]

Use of our expressions experimentally requires as a prerequisite measurements
of three parameters: the characterisitic impedance $Z_{0}$ of the
LL and the (boundary) interface resistances $Z_{S}$ and $Z_{D}$.
This can be readily done by $kHz$ AC conductance measurements as
explained in \cite{p3}.

For matching impedances we observe finally that the tunable impedances
need not be at the mesoscopic scale. This actually depends on the
measurement one is interested in: our calculations assume interface
resistances; usually the length over which there will be relaxation
is simply the inelastic scattering length $l_{inel}$. So if one works
at finite frequency $\omega$, one requires $\frac{v_{F}}{\omega}\gg l_{inel}$.
If the tunable impedances have a size smaller than $\frac{v_{F}}{\omega}$
then they can be considered as being interfacial (as assumed in our
calculations) and one need not worry over the spatial extent of the
contacting electrode. Therefore for matching impedances at the DC
level, the interface impedances can even be macroscopic. Only for
higher frequency measurements would one need mesoscopic contacts.

\appendix

\section{On fractionalization.}

\subsection{Motivation for the 'fractional states picture' of the Luttinger liquid.}

Exact solutions of models which belong to the Luttinger liquid universality
class do show fractional excitations. to cite but a few:

- the Heisenberg spin chain has a continuum of spin $1/2$ spinons
which are unaccounted for in the low energy mapping of a Luttinger
liquid (this would imply charge $1/2$ states in the Fock space of
the Jordan-Wigner fermions);

- the Hubbard model has also spinons but additionally shows charge
$+e$ spinless states, the holons.

\noindent Yet the description of the Luttinger liquid in the bosonization
scheme does not show any fractional states but reveals two kinds of
collective excitations are derived:

- plasmons (collective density fluctuations, the bosons of 'bosonization');

- 'zero-mode' operators which change the number of fermions by integral
increments (the density of left or right moving fermions is changed
uniformly: hence the name 'zero-mode'). 

(A note on terminology: by fractional states we mean states coming
from the fragmentation of the electron and which carry parts of the
quantum numbers of the original particle; we do not assume that these
parts are rational numbers. They might be irrational numbers.)

The solution of the apparent paradox is simple: the plasmon + zero
mode states constitute indeed a complete basis of eigenstates of the
LL and there can be no missing states in the diagonalization of the
LL. If fractional states exist in the LL they can only form as states
in alternate complete bases of eigenstates. This is what we actually
proved in \cite{fractionalization}.

Depending on the physical process under scrutiny a specific eigenbasis
may prove more or less convenient. One main drawback of the plasmon
+ zero-mode basis is that it is not fitted to describe the charge
dynamics in terms of \textbf{elementary processes} (involving diffusion
of few elementary excitations) because plasmons carry momentum but
\textbf{no} charge, while zero-mode excitations have charge but no
momentum. Describing the scattering of two fermions by a potential
using zero-modes and plasmons would involve an infinite number of
plasmon states (this follows from the fact that the fermion operator
is an exponential of plasmon operators).

Likewise it is not possible to interpret the shot noise results for
a LL with an impurity \cite{kane} or with a STM tip \cite{stm} in
terms of elementary processes using zero-mode and plasmon states.
Fractional states are mandatory. The natural language for transport
in a LL is that of fractional states for a LL and this shows up in
simpler calculations as this paper shows. So mathematically they are
useful tools.

\subsection{Fractional 'zero-mode' operators?}

While fractional states appear naturally in a field-theoretical approach
of bosonization they translate in the constructive approach 'a la
Heidenreich \& Haldane' \cite{bosonisation} into zero-mode operators
with a non-integral power such as:

\[
exp\, iK\theta_{0}\]

where $\theta_{0}$ is the usual (superfluid) phase conjugate to the
number operator $\widehat{N}$:\[
\left[\widehat{N},\theta_{0}\right]=-i\]
and with: \[
\left[\widehat{N},exp\, iK\theta_{0}\right]=K\: exp\, iK\theta_{0}\]
which shows that the states are fractional. This can raise consistency
problems due to the following technicality: if an operator has integer
eigenvalues then it is hermitian only in the space of states periodic
for its canonical conjugate field (see especially Appendix D.2 of
the second reference in \cite{bosonisation}). The above canonical
commutation relation for the number operator is therefore somewhat
an abuse of language since it would imply on the one hand $\left\langle N\left|\left[\widehat{N},\theta_{0}\right]\right|N\right\rangle =\left\langle N\left|-i\right|N\right\rangle =-i$
and on the other, by (erroneously) using the hermiticity of the operator
$\widehat{N}$: $\left\langle N\left|\widehat{N}\theta_{0}-\theta_{0}\widehat{N}\right|N\right\rangle =(N-N)\left\langle N\left|\theta_{0}\right|N\right\rangle =0$.
So one should apply $\widehat{N}$ to periodic functions of the phase
field $\theta_{0}$ such as $exp\, i\theta_{0}$and rather use the
commutation relation: \[
\left[\widehat{N},exp\, i\theta_{0}\right]=exp\, i\theta_{0}.\]

This implies that operators such as $exp\, iK\theta_{0}$ may lead
to similar hermiticity issues. 

Actually this difficulty is at the core of the explanation of how
fractional states can exist in a system made out of integral charges
(electrons) and \emph{in spite} of the difficulty indeed. 

As is clear from $\left[\widehat{N},exp\, iK\theta_{0}\right]=K\: exp\, iK\theta_{0}$
the operator $exp\, iK\theta_{0}$ has a zero expectation value between
states $\left\langle M\right|exp\, iK\theta_{0}\left|N\right\rangle =0$:
this is precisely stating that it has a non-integer charge. One might
imagine several ways out to ensure such an operator is properly defined:
one is to enlarge the Fock space to accomodate states such as $\left|N+K\right\rangle $
but this is of course forbidden. The structure of the Fock space is
rigid: it is defined by the electrons. The other way is to create
such a fractional state along with another one so that the total charge
is an integer: this is what precisely happens in our construction
of the fractional bases of states in a LL.

A fractional state is never created in isolation so that one never
has inconsistent expectation values; one only meets expressions of
the form:\[
\left\langle M\left|exp\, iQ_{1}\theta_{0}\: exp\, iQ_{2}\theta_{0}\right|N\right\rangle =\delta_{M,N+Q_{1}+Q_{2}}\]
where $Q_{1}+Q_{2}$ is an integral number. There is therefore no
hermiticity problem. All the expectation values involving these (pairs)
of fractional states are perfectly well defined. 

The constraint $Q_{1}+Q_{2}\in Z$ can be viewed as a selection rule
on the allowed fractional states. In the case of the LL the two fractional
states although created together need only obey such a selection rule
and apart from it are completely independent: they will generate continua
of excitations parametrized by the two independent dispersions of
the two fractional states.

\subsection{Description of the fractional states of the LL.}

(What follows is merely a heuristic description of the fractional
states. For details the reader is referred to Ref. \cite{fractionalization}).

The previous picture explaining how fractional states may arise is
a little bit more complicated with the actual LL because one then
has two species of electrons (left or right moving electrons) and
the low-energy Fock space is a direct product of the Fock spaces of
each fermion species. This means that one has two kinds of basic number
operators and associated canonical conjugate zero-mode fields:

\begin{eqnarray*}
\left[\widehat{N}_{R},\theta_{R,0}\right] & = & -i\\
\left[\widehat{N}_{L},\theta_{L,0}\right] & = & -i\end{eqnarray*}

This is not a big complication and one finds additional selection
rules. Let us see how.

It is convenient to consider the following zero modes:\begin{eqnarray*}
\left[\widehat{N},\theta_{0}\right] & = & -i\\
\left[\widehat{J},\phi_{0}\right] & = & -i\end{eqnarray*}
where: $\widehat{N}=\widehat{N}_{R}+\widehat{N}_{L}$ (the total charge)
and: $\widehat{J}=\widehat{N}_{R}-\widehat{N}_{L}$(which is related
to the current); obviously: $\theta_{0}=\frac{\theta_{R}+\theta_{L}}{2}$
and $\phi_{0}=\frac{\theta_{R}-\theta_{L}}{2}$. The fractional states
can be shown to involve operators such as:\[
exp\, iQ\left(\theta_{0}\pm\frac{\phi_{0}}{K}\right).\]
 Such combinations arise because the diagonalisation of the LL hamiltonian
involve similar combinations for the boson fields.

In order to have meaningful expectation values such as:

\[
\left\langle M_{R},M_{L}\left|exp\, iQ_{+}\left(\theta_{0}-\frac{\phi_{0}}{K}\right)\: exp\, iQ_{-}\left(\theta_{0}+\frac{\phi_{0}}{K}\right)\right|N_{R},N_{L}\right\rangle \]
it is straightforward to show one has to impose the following conditions
involving the integers $Q_{R}=M_{R}-N_{R}$ and $Q_{L}=M_{L}-N_{L}$: 

\[
\left(\begin{array}{c}
Q_{R}\\
Q_{L}\end{array}\right)=\left(\begin{array}{cc}
\frac{1+K^{-1}}{2} & \frac{1-K^{-1}}{2}\\
\frac{1-K^{-1}}{2} & \frac{1+K^{-1}}{2}\end{array}\right)\left(\begin{array}{c}
Q_{+}\\
Q_{-}\end{array}\right).\]
If we invert the relation we have the following constraints on the
fractional charges \[
\left(\begin{array}{c}
Q_{+}\\
Q_{-}\end{array}\right)=\left(\begin{array}{cc}
\frac{1+K}{2} & \frac{1-K}{2}\\
\frac{1-K}{2} & \frac{1+K}{2}\end{array}\right)\left(\begin{array}{c}
Q_{R}\\
Q_{L}\end{array}\right).\]
 (The index $\pm$ refers to two counterpropagating chiral modes of
the LL.)

Since $Q_{R}$ and $Q_{L}$ are arbitrary (positive or negative) integers
the spectrum of the allowed fractional charges forms a two-dimensional
\emph{lattice.}

As for any lattice all the states are spanned by \emph{primitive vectors}:
these vectors represent fractional excitations from which all the
others can be built; in other words they are \emph{elementary excitations}.
Here obviously for instance:\[
\left(\begin{array}{c}
Q_{+}\\
Q_{-}\end{array}\right)=Q_{R}\left(\begin{array}{c}
\frac{1+K}{2}\\
\frac{1-K}{2}\end{array}\right)+Q_{L}\left(\begin{array}{c}
\frac{1-K}{2}\\
\frac{1+K}{2}\end{array}\right)\]

But as for any lattice again the choice of a primitive basis is not
unique and therefore one will have several equivalent sets of elementary
fractional excitations. 

As explained above depending on the physical process scrutinized one
or another of these basis will be more adapted: in general it will
be better to use a basis involving the fewer number of elementary
excitations in order to truly describe elementary processes (involving
few particles).

The previous basis of states is convenient for processes involving
only one of the two species of fermions so that $Q_{R}=0$ or $Q_{L}=0$:
it involves two fractional states with charges $Q=\frac{1\pm K}{2}$.
These are precisely the states we considered in this paper for the
shot noise created by injection of particles by a STM.

Another basis is more convenient when one deals with particle-hole
excitations ($Q_{R}=-Q_{L}$ so that $Q=0$):\[
\left(\begin{array}{c}
Q_{+}\\
Q_{-}\end{array}\right)=Q\left(\begin{array}{c}
\frac{1-K}{2}\\
\frac{1+K}{2}\end{array}\right)+\frac{Q+J}{2}\left(\begin{array}{c}
K\\
-K\end{array}\right);\]
for $Q_{R}=-Q_{L}$, $J=2Q_{R}$ is an even integer and the equation
simplifies into:\[
\left(\begin{array}{c}
Q_{+}\\
Q_{-}\end{array}\right)=\frac{J}{2}\left(\begin{array}{c}
K\\
-K\end{array}\right).\]
These excitations with charge $K$ are actually the analogs of the
Laughlin quasiparticles of the Fractional Qunatum Hall Effect. Indeed
for $K=\frac{1}{2n+1}$ the LL hamiltonian is identical to two counterpropagating
copies of the chiral Luttinger liquid edge states at filling $\nu=\frac{1}{2n+1}$;
the operator for the Laughlin quasiparticle of the edge states then
coincides exactly with the fractional operator of charge $K$ considered
here. The main difference is that the charge $K$ need not be a rational
number.

Another convenient basis involves holon states. If one generalizes
these considerations to a spinful LL one finds that such a spinless
state is created along with a spin $\frac{1}{2}$ chargeless state,
the spinon. One thus recovers the fractional excitations of the Hubbar
model. As an aside we mention that the holon state is actually \emph{dual}
to the charge $K$ Laughlin quasiparticle (i.e. electromagnetic duality
- which exchanges the roles of the electric and magnetic field, or
here for the LL, which exchanges current and charge - maps the holon
state on the Laughlin quasiparticle). There is therefore a deep connection
between the holon of the Hubbard model and the Laughlin quasiparticle
of the Fractional Quantum Hall Effect, which should probably come
as a surprise.

Finally we mention that for bosons (and spins) different selection
rules must be used: although the low-energy field theory (the LL hamiltonian)
is the same as that of fermions there are still remnants of the exchange
statistics. One finds that: \[
\left(\begin{array}{c}
Q_{+}\\
Q_{-}\end{array}\right)=Q\left(\begin{array}{c}
\frac{1}{2}\\
\frac{1}{2}\end{array}\right)+J\left(\begin{array}{c}
K\\
-K\end{array}\right);\]
 there are two basic elementary excitations: one is the Laughlin quasiparticle,
the other is a charge $\frac{1}{2}$ state which simply corresponds
for spin systems to the spinon.

\begin{figure}[H]
\includegraphics{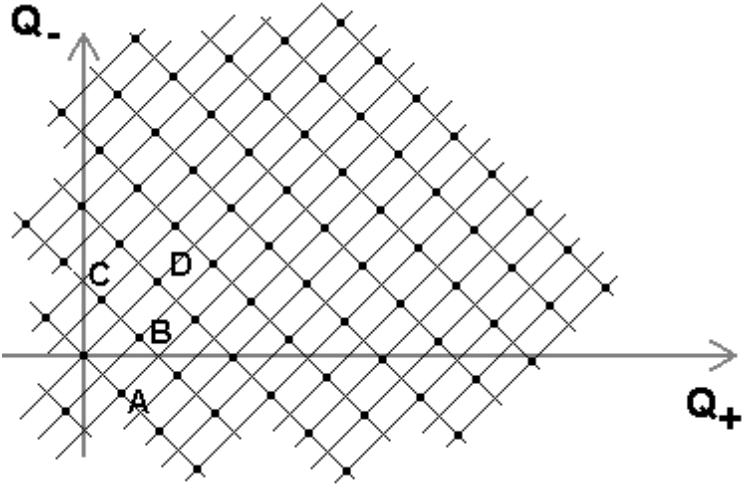}

\caption{\label{cap:fermion lattice}Fractional excitations for fermions:
the allowed states are the nodes of a rectangular centred Bravais
lattice. Point 'A' corresponds to a Laughlin quasiparticle-quasihole
pair; 'B' (resp. 'C') corresponds to the creation of a pair with charge
$\frac{1+K}{2}$ and $\frac{1-K}{2}$ (resp. $\frac{1-K}{2}$ and
$\frac{1+K}{2}$) ; 'D' is a holon (or more aptly a chargeon since
it is negatively charged) state.}
\end{figure}

\begin{figure}[H]
\includegraphics{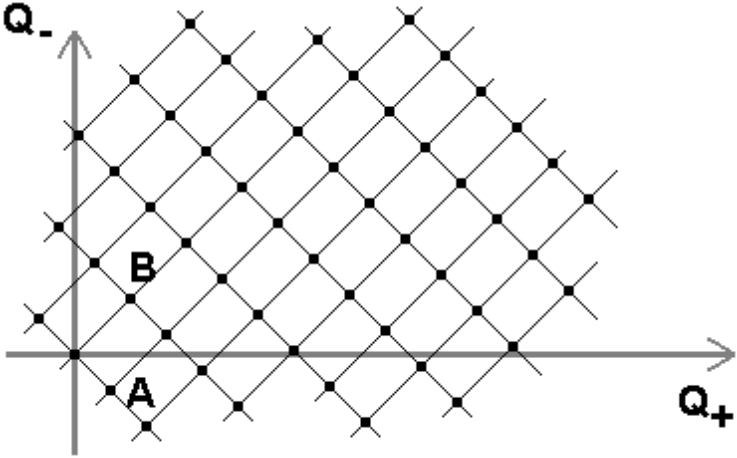}

\caption{\label{cap:boson lattice}Fractional excitations for bosons: the
allowed states are the nodes of a rectangular Bravais lattice. Point
'A' corresponds to a Laughlin quasiparticle-quasihole pair; 'B' corresponds
to the creation of a pair with charge $\frac{1}{2}$ and $\frac{1}{2}$
(interpreted as spinon states for spin systems when transformed into
hard-core bosons).}
\end{figure}

The lattice for fermions is a rectangular centred lattice whose axes
are the directions $Q_{+}\pm Q_{-}=0$, while for bosons it is a rectangular
lattice (with similar axes). These lattices become square lattice
for the special values $K=1$ (fermions) or $K=1/2$ (bosons or spins),
which are actually self-dual points in terms of the electromagnetic
duality discussed above. For these values the spectrum of elementary
excitations involves only one kind of elementary excitation: the free
fermion on the one hand, and the spinon on the other hand ($K=1/2$
corresponds to the $SU(2)$ symmetric Heisenberg spin chain).

\section{Green's functions using {}``impedance boundary conditions'': towards
a keldysh treatment\label{sec:Green's-functions-using}.}

We define the standard phase field of the LL boson Hamiltonian per:\[
\rho(x,t)-\rho_{0}=\frac{1}{\sqrt{\pi}}\partial_{x}\Phi.\]

Using that definition, given the reflection coefficients for the density
$r_{S}=\frac{Z_{S}-Z_{0}}{Z_{S}+Z_{0}}$ and $r_{D}=\frac{Z_{D}-Z_{0}}{Z_{D}+Z_{0}}$
at the source and drain boundaries, the reflection coefficients for
the phase field $\Phi$ are easily shown to be $-r_{S}$ and $-r_{D}$,
namely the chiral components $\Phi_{+}(x,t)=\Phi_{+}(x-ut,0)$ and
$\Phi_{-}(x,t)=\Phi_{-}(x+ut,0)$ obey the boundary conditions:\begin{eqnarray*}
\Phi_{+}(-L/2,t) & = & -r_{S}\,\Phi_{-}(-L/2,t),\\
\Phi_{-}(L/2,t) & = & -r_{D}\,\Phi_{+}(L/2,t).\end{eqnarray*}
Due to these one can view the propagation as being in a doubled length
system (in a loop).

The Green's function $G(x,t;y,0)$ is derived in a standard manner
by solving its equation of motion:\[
\left[\frac{1}{u^{2}}\frac{\partial^{2}}{\partial t^{2}}+\frac{\partial^{2}}{\partial x^{2}}\right]G(x,t;y,0)=\frac{K}{u}\delta(x-y)\delta(t).\]

The Green's function can be conveniently divided into four chiral
components: $G=G^{++}+G^{--}+G^{+-}+G^{-+}$ where $G^{\pm\pm}=-i\left\langle T\Phi_{\pm}\Phi_{\pm}\right\rangle $
for the causal Green's functions and appropriate definitions for the
retarded Green's function and so forth. After Fourier transforming
one seeks a solution of the following form:

\[
G^{\pm\pm}(x,y,\omega)=\left[\theta(x-y)\, a^{\pm}(\omega)+\theta(y-x)\, b^{\pm}(\omega)\right]\, e^{\pm i\frac{\omega}{u}(x-y)}\]
because $G^{\pm\pm}(x,y,t)$ obey respectively the equations of motion:
$\left[\frac{\pm1}{u}\frac{\partial}{\partial t}+\frac{\partial}{\partial x}\right]G^{\pm\pm}(x,t;y,0)=\frac{K}{2u}\delta(x-y)\delta(t)$.

We enforce the 'impedance boundary conditions' which imply that:\begin{eqnarray*}
b^{+} & = & r_{S}r_{D}\, e^{i2\frac{\omega}{u}L}a^{+}\\
a^{-} & = & r_{S}r_{D}\, e^{i2\frac{\omega}{u}L}b^{-}\\
G^{+-}(x,y,\omega) & = & -r_{S}G^{--}(-L-x,y,\omega)\\
 & = & -r_{S}\, b^{-}\, e^{i\frac{\omega}{u}(x+y+L)}\\
G^{-+}(x,y,\omega) & = & -r_{D}G^{++}(x,L-y,\omega)\\
 & = & -r_{D}\, b^{-}\, e^{i\frac{\omega}{u}(-x-y+L)}\end{eqnarray*}

Finally the equations of motion for the chiral Green's functions are
used to extract the undetermined coefficients so that:\begin{eqnarray*}
G(x,y,\omega) & = & \frac{K}{2i\omega}\frac{1}{1-r_{S}r_{D}e^{i2\phi}}\\
\{ & \theta(x-y) & \left[e^{i\frac{\omega}{u}(x-y)}+r_{S}r_{D}e^{-i\frac{\omega}{u}(x-y-2L)}\right]\\
+ & \theta(y-x) & \left[e^{-i\frac{\omega}{u}(x-y)}+r_{S}r_{D}e^{i\frac{\omega}{u}(x-y+2L)}\right]\\
- & r_{S}\, e^{i\frac{\omega}{u}(x+y+L)}-r_{D}\, e^{i\frac{\omega}{u}(-x-y+L)} & \}\end{eqnarray*}
where $\phi=\frac{\omega L}{u}$ and where the poles are shifted from
the real axis according to the usual prescriptions for the causal
or retarded Green's functions, etc. The interpretation of the Green's
function is quite straightforward: to go from one point to the other
there are four kinds of basic trajectories, (i) if one is behind the
destination going straight to it, or (ii-iii) going after bouncing
from one boundary or the other, and (iv) lastly going after bouncing
two times from different boundaries. These basic trajectories must
then be convoluted by round trips along the whole loop (of length
$2L$) which yield the overall factor $\left(1-r_{S}r_{D}e^{i2\phi}\right)^{-1}$.

The Green's function coincides with that of the infinite LL when impedance
matching is realized namely: $Z_{0}=Z_{S}=Z_{D}$ which in turn implies:
$r_{S}=r_{D}=0$.

The basic ingredient to use the Keldysh formalism is the Keldysh Green
function matrix for that field $\Phi$. The upper time-line is indexed
by $+$ while the lower time-reversed line is indexed by sign $-$.
The following correlator can be extracted from the Green's function
as :\begin{eqnarray*}
F_{-+}(x,y,t)= & \left\langle \Phi(x,t)\Phi(y,0)\right\rangle \\
= & -\frac{K}{4\pi}\;\sum_{n=-\infty}^{+\infty}\left(r_{S}r_{D}\right)^{\left|n\right|}\\
\{ & \ln\left[\delta+i(ut+2nL)\right]+(x-y)^{2}\\
 & -r_{S}\:\ln\left[\delta+i(ut+x+y+(2n+1)L)\right]\\
 & -r_{D}\:\ln\left[\delta+i(ut-x-y+(2n+1)L)\right] & \}\end{eqnarray*}
The successive factors correspond to either direct propagation or
propagation after reflections at the boundaries. The other matrix
elements follow immediately through their definitions $F_{+-}(x,y,t)=\left\langle \Phi(y,0)\Phi(x,t)\right\rangle =F_{-+}(y,x,-t)$
and likewise $F_{++}(x,y,t)=\theta(t)F_{-+}(x,y,t)+\theta(-t)F_{+-}(x,y,t)$,
$F_{--}(x,y,t)=\theta(t)F_{+-}(x,y,t)+\theta(-t)F_{-+}(x,y,t)$.

The inhomogeneous LL is found again to be a special case of our general
expressions: $r_{S}=r_{D}=-\gamma=-\frac{1-K}{1+K}$.

Starting from these one can then use the general relations derived
in \cite{stm} giving the noise spectrum as a function of the Keldysh
Green functions: these relations are valid quite generally since resulting
from perturbation theory and do not depend on the use of the inhomogeneous
LL model.

\section{Injected vs measured currents for AC transport. }

Let us prove Eq.(\ref{ac-ren}). We must modify the equations defining
the currents by specifying the position.

\begin{align*}
I_{2}(t) & =I_{R}(L_{2},t)=i_{R}^{+}(L_{2},t)-i_{R}^{-}(L_{2},t)\\
-I_{1}(t) & =I_{L}(-L_{1},t)=i_{L}^{+}(-L_{1},t)-i_{L}^{-}(-L_{1},t)\end{align*}
\begin{align*}
I_{T}^{+}(t) & =i_{R}^{+}(0,t)-i_{L}^{+}(0,t)\\
I_{T}^{-}(t) & =i_{L}^{-}(0,t)-i_{R}^{-}(0,t)\end{align*}
 And the boundary conditions:

\begin{align*}
I_{1} & =\frac{Z_{0}}{Z_{S}}\left[i_{L}^{+}(-L_{1},t)+i_{L}^{-}(-L_{1},t)\right]\\
I_{2} & =\frac{Z_{0}}{Z_{D}}\left[i_{R}^{+}(L_{2},t)+i_{R}^{-}(L_{2},t)\right]\end{align*}
 We can rewrite eveything in terms of fields at the position $x=0$
by taking into account the fact that the fields being chiral: \[
i_{R}^{\pm}(L_{2},t)=e^{\pm i\phi_{1/2}}~i_{R}^{\pm}(0,t),\quad i_{L}^{\pm}(-L_{1},t)=e^{\mp i\phi_{1/2}}~i_{L}^{\pm}(0,t)\]
 It is convenient to define the vectors: \[
\overrightarrow{i_{R}}=\left(\begin{array}{c}
i_{R}^{+}(0)\\
i_{R}^{-}(0)\end{array}\right),\quad\overrightarrow{i_{L}}=\left(\begin{array}{c}
i_{L}^{+}(0)\\
i_{L}^{-}(0)\end{array}\right).\]
 Then the boundary conditions can be recast as: \begin{align*}
Z_{S}I_{1} & =Z_{0}\left(\begin{array}{cc}
\exp-i\phi_{1}, & \exp i\phi_{1}\end{array}\right)\cdot\overrightarrow{i_{L}}\\
Z_{D}I_{1} & =Z_{0}\left(\begin{array}{cc}
\exp i\phi_{2}, & \exp-i\phi_{2}\end{array}\right)\cdot\overrightarrow{i_{R}}\end{align*}
 while the definition of the currents imply: \begin{align*}
I_{1} & =\left(\begin{array}{cc}
-\exp-i\phi_{1}, & \exp i\phi_{1}\end{array}\right)\cdot\overrightarrow{i_{L}}\\
I_{2} & =\left(\begin{array}{cc}
\exp i\phi_{2}, & -\exp-i\phi_{2}\end{array}\right)\cdot\overrightarrow{i_{R}}\end{align*}
 Or in a matrix form: \begin{align*}
I_{1}\left(\begin{array}{c}
Z_{S}\\
1\end{array}\right) & =\left(\begin{array}{cc}
Z_{0}\exp-i\phi_{1} & Z_{0}\exp i\phi_{1}\\
-\exp-i\phi_{1} & \exp i\phi_{1}\end{array}\right)\overrightarrow{i_{L}}\\
I_{2}\left(\begin{array}{c}
Z_{D}\\
1\end{array}\right) & =\left(\begin{array}{cc}
Z_{0}\exp i\phi_{2} & Z_{0}\exp-i\phi_{2}\\
\exp-i\phi_{2} & -\exp-i\phi_{2}\end{array}\right)\overrightarrow{i_{R}}\end{align*}
 We solve for $\overrightarrow{i_{L}}$ and $\overrightarrow{i_{R}}$:
\[
\overrightarrow{i_{L}}=\frac{I_{1}}{2Z_{0}}\left(\begin{array}{c}
\left(Z_{S}-Z_{0}\right)\exp i\phi_{1}\\
\left(Z_{S}+Z_{0}\right)\exp-i\phi_{1}\end{array}\right),\quad\overrightarrow{i_{R}}=\frac{I_{2}}{2Z_{0}}\left(\begin{array}{c}
\left(Z_{D}+Z_{0}\right)\exp-i\phi_{2}\\
\left(Z_{D}-Z_{0}\right)\exp i\phi_{2}\end{array}\right)\]
 The tunneling currents at position $x=0$ and frequency $\omega$
are thus: \begin{align*}
\left(\begin{array}{c}
I_{T}^{-}\\
I_{T}^{+}\end{array}\right) & =\left(\begin{array}{cc}
0 & -1\\
1 & 0\end{array}\right)\left(\overrightarrow{i_{R}}-\overrightarrow{i_{L}}\right)=\left(\begin{array}{cc}
0 & -1\\
1 & 0\end{array}\right)\frac{1}{2Z_{0}}\left(\begin{array}{c}
-I_{1}\left(Z_{S}-Z_{0}\right)\exp i\phi_{1}+I_{2}\left(Z_{D}+Z_{0}\right)\exp-i\phi_{2}\\
-I_{1}\left(Z_{S}+Z_{0}\right)\exp-i\phi_{1}+I_{2}\left(Z_{D}-Z_{0}\right)\exp i\phi_{2}\end{array}\right)\\
 & =\frac{1}{2Z_{0}}\left(\begin{array}{cc}
\left(Z_{S}+Z_{0}\right)\exp-i\phi_{1} & -\left(Z_{D}-Z_{0}\right)\exp i\phi_{2}\\
-\left(Z_{S}-Z_{0}\right)\exp i\phi_{1} & \left(Z_{D}+Z_{0}\right)\exp-i\phi_{2}\end{array}\right)\left(\begin{array}{c}
I_{1}\\
I_{2}\end{array}\right)\end{align*}
 Inverting the matrix yields: \begin{align*}
\left(\begin{array}{c}
I_{1}\\
I_{2}\end{array}\right) & =\frac{2Z_{0}}{\left(Z_{S}+Z_{0}\right)\left(Z_{D}+Z_{0}\right)\exp-i\left(\phi_{1}+\phi_{2}\right)-\left(Z_{S}-Z_{0}\right)\left(Z_{D}-Z_{0}\right)\exp i\left(\phi_{1}+\phi_{2}\right)}\\
 & \times\left(\begin{array}{cc}
\left(Z_{D}+Z_{0}\right)e^{-i\phi_{2}} & \left(Z_{D}-Z_{0}\right)e^{i\phi_{2}}\\
\left(Z_{S}-Z_{0}\right)e^{i\phi_{1}} & \left(Z_{S}+Z_{0}\right)e^{-i\phi_{1}}\end{array}\right)\left(\begin{array}{c}
I_{T}^{-}\\
I_{T}^{+}\end{array}\right)\end{align*}
 where $\left(\begin{array}{c}
I_{T}^{-}\\
I_{T}^{+}\end{array}\right)$ are the currents injected at position $x=0$ (the STM tip). Finally
we note that the currents injected at $x=0$ get a phase dependence
when reaching the boundaries so that: $\left(\begin{array}{c}
I_{T}^{-}(-L_{1},\omega)\\
I_{T}^{+}(L_{2},\omega)\end{array}\right)=\left(\begin{array}{c}
e^{i\phi_{1}}I_{T}^{-}\\
e^{i\phi_{2}}I_{T}^{+}\end{array}\right)$ (this follows simply from the chirality of these currents: $f(t,x)=f(t-x/c,0)$
implies of course that $f(\omega,x)=e^{i\omega x/c}f(\omega,0)$ ).
Replacing these expressions we get Eq.(\ref{ac-ren}).


\begin{thebibliography}{10}
\bibitem{noise}Ya. M. Blanter, M. Büttiker, Phys. Rep. \textbf{336}, 1 (2000).
\bibitem{fqhe}R. de Picciotto et al, Nature (London) \textbf{389}, 162 (1997); L.
Saminadayar, D.C. Glattli, Y. Jin, B. Etienne, Phys. Rev. Lett. \textbf{79},
2526 (1997).
\bibitem{ll}F. D. M. Haldane, J. Phys. C \textbf{14}, 2585 (1981); H.J. Schulz,
in Les Houches Summer School 1994, Mesoscopic Quantum Physics, E.
Akkermans et al. eds, Elsevier Science, Amsterdam (1995).
\bibitem{bosonisation}R. Heidenreich, R. Seiler, A. Uhlenbrock, J. Stat. Phys. \textbf{22},
27 (1980); J. von Delft, H. Schoeller, Annalen Phys. \textbf{7}, 225
(1998).
\bibitem{fractionalization}K.-V. Pham, M. Gabay, P. Lederer, Phys. Rev. B \textbf{61}, 16397
(2000).
\bibitem{kane}C. L. Kane and M. P. A. Fisher, Phys. Rev. Lett. \textbf{72}, 724
(1994).
\bibitem{ill}D. Maslov, M. Stone, Phys. Rev. B \textbf{52}, R5539 (1995); I. Safi,
H. Schulz, Phys. Rev. B \textbf{52}, R17040 (1995); V. V. Ponomarenko,
Phys. Rev. B \textbf{52}, R8666 (1995).
\bibitem{stm}A. Crépieux, R. Guyon, P. Devillard, T. Martin, Phys. Rev. B. \textbf{6}7,
205408 (2003); A. Lebedev, A. Crépieux, T. Martin, cond-mat/0405325.
\bibitem{pono}V. V. Ponomarenko and N. Nagaosa, Solid State Commun. 110, 321 (1999).
\bibitem{impurity}B. Trauzettel, F. Dolcini, I. Safi, H. Grabert, Phys. Rev. Lett. \textbf{92},
226405 (2004); F. Dolcini, B. Trauzettel, I. Safi, H. Grabert, cond-mat/0409320.
\bibitem{p3}K-V Pham, Eur. Journ. Phys. B 36, 607 (2003).
\bibitem{lc}M. W. Bockrath, Ph. D. Thesis, University of California, Berkeley,
(1999); P. J .Burke, IEEE Trans. Nanotechn. 1, 129 (2002); IEEE Trans.
Nanotechn. 2, 155 (2003).
\bibitem{p1}K-I Imura, K-V Pham, F. Pi\'{e}chon, P. Lederer, Phys. Rev. B. \textbf{66},
035313 (2002).
\bibitem{p2}K-V Pham, F. Pi\'{e}chon, K-I Imura, P. Lederer, Phys. Rev. B. \textbf{68},
205110 (2003).
\bibitem{kcond}C. L. Kane and M. P. A. Fisher, Phys. Rev. B. 46, 15233 (1992).
\bibitem{key-5}C. L. Kane and M. P. A. Fisher, in \emph{Perspectives in the Quantum
Hall Effects}, Eds S. Das Sarma, A. Pinczuk (Wiley, New York, 1995).
\bibitem{key-6}R. Egger, H. Grabert, Phys. Rev. B \textbf{58}, 10761 (1998).
\bibitem{key-3}I. Safi, Ann. Phys. Fr. \textbf{22}, 463 (1997).
\bibitem{chamon}C. de Chamon, D. E. Freed, X. G. Wen, Phys. Rev. B \textbf{53}, 4033
(1996).\end{thebibliography}
\end{document}